\documentclass{article}

\usepackage{arxiv}
\usepackage[OT2, T1]{fontenc}
\usepackage[russian, english]{babel}
\usepackage{caption}
\usepackage{lscape} 
\usepackage{amsmath}
\usepackage{geometry}
\usepackage{amsthm}
\usepackage{adjustbox}
\usepackage{amsfonts}
\usepackage{mathtools}
\usepackage{verbatim}
\usepackage{algpseudocode,algorithm,algorithmicx}
\usepackage{mathtools}
\usepackage{comment}
\usepackage{verbatim, times, booktabs, dcolumn, setspace, fullpage}

\newtheorem{theorem}{Theorem}

\newtheorem{remark}{Remark}
\usepackage{graphicx}
\usepackage{wrapfig}
\usepackage{caption}
\usepackage{mathtools}
\usepackage{float}
\usepackage{multirow}

\usepackage{makecell}

\allowdisplaybreaks

\usepackage[english,noautotitles-r]{SASnRdisplay} 
\usepackage{url}
\usepackage{subfigure}
\lstdefinestyle{r-output}{
style = r-style,
style = r-output-user,
}

\usepackage{url}            
\usepackage{booktabs}       
\usepackage{amsfonts}       
\usepackage{nicefrac}       
\usepackage{microtype}      
\usepackage{graphicx}
\usepackage{natbib}
\usepackage{doi}

\newcommand{\E}{\mathbf{E}}

\title{Two-Sample Testing with Missing Data via Energy Distance: Weighting and Imputation Approaches}
\author{ \href{https://orcid.org/0000-0002-0460-400X}{\includegraphics[scale=0.5]{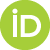}\hspace{1mm} Danijel G.    Aleksi\' c} \\
	University of Belgrade\\
	Faculty of Organizational Sciences, Faculty of Mathematics\\
	Belgrade, 11000, Serbia \\
	\texttt{danijel.aleksic@fon.bg.ac.rs} \\
	\And
	\href{https://orcid.org/0000-0001-8243-9794}{\includegraphics[scale=0.5]{OrcidLogo.eps}\hspace{1mm}Bojana Milo\v sevi\' c} \\
	University of Belgrade\\
        Faculty of Mathematics\\
	Belgrade, 11000, Serbia \\
	\texttt{bojana@matf.bg.ac.rs} \\
}
\date{}

\begin{document}
\maketitle

\begin{abstract}
In this paper, we address the problem of two-sample testing in the presence of missing data under a variety of missingness mechanisms. Our focus is on the well-known energy distance-based two-sample test. In addition to the standard complete-case approach, we propose a modification of the test statistic that incorporates all available data, utilizing appropriate weights. The asymptotic null distribution of the test statistic is derived and two resampling procedures for approximating the corresponding p-values are proposed. We also propose a new bootstrap method specifically designed for a test statistic based on samples completed via common imputation methods.
Through an extensive simulation study, we compare all methods in terms of type I error control and statistical power across a set of sample sizes, dimensions, distributions, missingness mechanisms, and missingness rates. Based on these results, we provide general recommendations for each considered scenario.

 \vspace{10pt}
\end{abstract}

\section{Introduction}

Testing whether two samples originate from the same probability distribution, known as two-sample testing, is a fundamental problem in statistical theory with a broad spectrum of applications across various fields. In medical research, such tests are used to compare patient outcomes between treatment and control groups, assessing the effectiveness of new drugs or interventions  \citep[see e.g.][]{pocock2013clinical}. In finance and economics, they help determine whether stock return distributions differ before and after major market events or whether income distributions vary across different regions. Additionally, in quality control, manufacturers employ these methods to verify whether changes in production processes impact product characteristics. Environmental scientists utilize them to assess changes in climate variables, such as temperature distributions over decades, while geneticists apply them to compare gene expression profiles between different populations \citep[see e.g.][]{borgwardt2006integrating}. 


Given the versatility of possible applications, two-sample tests continue to play a crucial role in data-driven decision-making across numerous disciplines, and, consequently, they have been extensively studied throughout the literature. Here, we present some notable examples. The idea of testing whether two samples originate from the same probability distribution first appeared for one-dimensional data, and can be traced back to \cite{pearson1900x}, who developed the well-known $\chi^2$-test. Later, in the early 20th century, \cite{student1908probable} tested for the difference between normal distributions by comparing the means; the idea was later expanded by \cite{fisher1925statistical} in his works on Analysis of Variance (ANOVA). These methods were parametric, and they had severe limitations in non-normal settings. 

First nonparametric solutions for this problem were presented in the  form of a well-known Kolmogorov-Smirnov (KS) test, which was developed by \cite{kolmogorov1933sulla} and extended by \cite{smirnov1939estimation}. The test was based on the measuring discrepancy of ECDFs between the two distributions. Wilcoxon–Mann–Whitney test \citep{wilcoxon1945individual, mann1947test} was another nonparametric two-sample test for one-dimensional data that was based on comparing the rank-sums between groups. Since then, many different univariate tests has been been proposed.


Test by \cite{hotelling1931generalization} can be considered to be one of the first two-sample tests for the multivariate data. The test was based on a test statistic that was generalizing Student's ratio, and, as such, it was reliant on the normality assumption, as well as the assumption of equal covariances. Notable example of a nonparametric multivariate two-sample test is the test by \cite{friedman1979multivariate} that generalized the two-sample runs test by \cite{wald1940test}. 

Some of the modern and commonly examples are the energy-distance-based two-sample test \citep{baringhaus2004new, szekely2004testing}, as well as the Maximum Mean Discrepancy (MMD) kernel-based approach by \cite{gretton2012kernel}. Their equivalence is shown by \cite{sejdinovic2013equivalence}.
Here, we study this problem in the presence of missing data under the broad class of missingness mechanisms. For this purpose, we focus on the energy-based form of the two-sample test \citep{baringhaus2004new, szekely2004testing}. Besides the complete-case approach, we propose a novel adaptation of the test statistic that utilizes all available data and explore its asymptotic properties, along with two resampling procedures for approximating the corresponding p-values. A novel bootstrap method is also introduced for p-value approximation when the test statistic is computed on samples filled using commonly used imputation methods. In an extensive simulation study, all approaches are compared in terms of preservation of type I error and in terms of power, under various sample sizes, dimensions, distributions, missingness mechanisms, and proportions, both on synthetic and real data. General recommendations are given for each of the studied scenarios. 

Results obtained from the real-world data indicate that complex data types like audio and images, despite their high dimensionality, tend to be structured around a lower-dimensional manifold. Due to this fact, the standard energy distance-based two-sample test, which is built upon Euclidean distance, performs well for low-dimensional data, and loses power quickly for data in high dimensional space \citep[e.g.][and references therein]{chu2024manifold}. Consequently, much work has been done in that direction: to modify the original test to be able to recognize the inner structure of the data and effectively reduce the dimension. A nice overview can be found in a paper by \cite{chu2024manifold}. However, in terms of adapting the original test to work with incomplete samples, not much progress has been made. Knowing the importance and the wide applicability of the test, it is essential to address that issue.  

Obtained results are presented according to the following structure. In Section \ref{energy:energy_test}, we restate some basic properties of the original energy test necessary for our further research. In Section \ref{energy:novel_procedures}, we introduce our novel testing procedures and outline their expected strengths and flaws. Section \ref{energy:power_study} is devoted to the extensive simulation study on synthetic data, with the aim to examine the performance of novel procedures in terms of preservation of type I error and in terms of empirical power.  In Section \ref{power_study_real} we extend the previous empriical study to the real data.
Depending on the specific scenario, such as data distribution and the underlying missingness mechanism, certain recommendations are provided at the end. Concluding remarks and potential extensions of this work are presented in Section \ref{sec:reflections}.

\section{Revisiting the energy test}\label{energy:energy_test}

Let $\mathcal{M}$ be a metric space and let $\mu$ and $\nu$ be two probability measures on it. Assume we have two independent samples of independent identically distributed (IID) random elements in $\mathcal{M}$: 
\begin{align}\label{samples}
    X_1, \dots, X_n \sim \mu \quad \text{and} \quad Y_1, \dots, Y_m \sim \nu.
\end{align}
A commonly tested hypothesis in a two-sample test is whether the two samples originate from the same distribution
\begin{align}\label{two_sample_H0}
    H_0: \, \mu = \nu,
\end{align}
against the complementary alternative. If $\rho$ is a metric on $\mathcal{M}$, one can define the \emph{energy distance} between the distributions $\mu$ and $\nu$ as
\begin{align}\label{energy_distance_def}
    D(\mu, \nu) = 2\E \rho (X, Y) - \E \rho (X, X') - \E \rho(Y, Y'),
\end{align}
where $X, X' \sim \mu$ and $Y, Y' \sim \nu$ are all mutually independent. 

The name \textit{energy distance} comes from physics. In physics, particularly in electrostatics, gravity, or molecular systems, we often deal with systems of many particles that interact pairwise (e.g., via gravitational or electric forces). In such a system, we can distinguish: (i) energy due to interactions within a group (e.g., within one cloud of particles), and (ii) energy due to interactions between particles from different groups. Physicists are often interested in the \textit{net interaction energy}, which is the difference between the total interaction energy and the internal self-energies. His net interaction energy tells us how much energy arises because these two systems are different, beyond what they internally contribute?

The term $\E \rho (X,Y)$ in \ref{energy_distance_def} can be seen as a total interaction energy between $\mu$ and $\nu$, i.e. the average distance between the two randomly chosen points from them. Similarly, $\E \rho (X, X')$ ($\E \rho (Y, Y')$) can be seen as an internal energy of $\mu$ ($\nu$),  the average distance between the two randomly chosen points from $\mu$ ($\nu$). So, when we are computing the energy distance $D(\mu, \nu)$, we are basically computing the \textit{excess interaction energy} between $\mu$ and $\nu$, beyond the energy one would expect if both samples were from the same distribution.

To further clarify, one could ask a question: Is the separation between the distributions more than what we would expect just due to their natural variability? If the answer is negative, we expect energy distance to be close to zero, and to significantly differ from zero otherwise.  For additional parallels to physics and further motivation, we recommend consulting the paper by \cite{szekely2013energy}.

The test statistic of the energy test \citep{baringhaus2004new, szekely2004testing} is based on the sample estimate of $D(\mu, \nu)$ and is defined as 
\begin{align}\label{energy_test_statistic}
    T_{nm} = \frac{2}{nm} \sum_{i=1}^n \sum_{j=1}^m \rho (X_i, Y_j) - \frac{1}{n^2} \sum_{i=1}^n \sum_{j=1}^n \rho (X_i, X_j) - \frac{1}{m^2} \sum_{i=1}^n \sum_{j=1}^n \rho (Y_i, Y_j).
\end{align}

The energy distance is not always the proper metric on the set of all probability distributions on $\mathcal{M}$ sometimes it is not positive; a sufficient condition, for example, is for $(\mathcal{M}, \rho)$ to be a metric space of a \emph{strong negative type}. Specifically, $(\mathcal{M}, \rho)$ is of a strong negative type if for any two probability measures $\mu$ and $\nu$, with finite first moments, on it it holds that  
\begin{align*}
    \iint \rho (u,v) \mathrm{d}\mu (u) \mathrm{d}\mu (v) + \iint \rho (u,v) \mathrm{d}\nu (u) \mathrm{d}\nu (v) - 2 \iint \rho (u,v) \mathrm{d}\mu (u) \mathrm{d}\nu (v) \leq 0,
\end{align*}
and the left hand side is equal to zero if and only if $\mu = \nu$. As a consequence, on a space of strong negative type, the energy distance is able to differentiate between any two probability distributions. An example of one such space, that is of key interest in mathematical statistics, is any separable Hilbert space. As a special case, we have the Euclidean space $(\mathbb{R}^d, \| \cdot \|)$ with the standard Euclidean metric, which is of our main interest. 
For more details, one should consult, e.g., \cite{klebanov2005n}, \cite{gonzalez2024consistent}, \cite{chu2024manifold}, and references therein.

\section{Novel procedures}\label{energy:novel_procedures}

In this section, we propose several novel adaptations of energy test suitable for incomplete samples with missingness mechanisms that are not necessarily MCAR. The main assumption is that samples \eqref{samples} are independent, and both consist of IID random vectors from $\mathbb{R}^d$. Generally speaking, our methodology can work with an arbitrary metric $\rho$ on $\mathbb{R}^d$, but our main focus will be on the standard Euclidean metric, which will be thoroughly examined in the extensive simulation study that will follow, where novel approaches will be compared to the complete case analysis used as a benchmark. 

First of all, let us introduce some basic notation that we will need. For every element $X_i$ of the sample $X_1, \dots, X_n$ let $R_i^X$ be the corresponding vector of the same length $d$ as $X_i$, whose $k$th element is equal to 1 if the $k$th element of $X_i$ is observed, and 0 otherwise. We will refer to $R_i^X$ as a \emph{response indicator vector} of $X_i$. Let $S_i^X$ be the indicator that each component of $X_i$ is observed, i.e. indicator that $X_i$ is the \emph{complete case}. It is readily seen that $S_i^X = \prod_{k=1}^n R_{i, k}^X$, where $R_{i, k}^X$ is the $k$th component of $R_i ^X$. Let $R_i^Y$ and $S_i^Y$ be defined in a similar manner.  

The number of complete cases from the sample $X_1, \dots, X_n$ will be denoted as $\hat{n}$; it is clear that $\hat{n} = \sum_{i=1}^n S_i^X$. Similarly, let $\hat{m} = \sum_{j=1}^m S_j^Y$ be the number of complete cases from the sample $Y_1, \dots, Y_m$. Finally, let $\odot$ denote the standard Hadamard--Schur componentwise multiplication:
\[ (u_1, u_2, \dots, u_d) \odot (v_1, v_2, \dots, v_d) = (u_1v_1, u_2v_2, \dots, u_dv_d). \]

\subsection{Complete-case analysis: the benchmark}

The first and fairly common approach in practice is complete-case analysis, where the test statistic $T_{nm}$ is calculated only on the fully observed sample elements. In our notation, the statistic can be written as
\begin{align}\label{T_CC}
    T_{nm}^{CC} =  \frac{2}{\hat{n} \hat{m}} \sum_{i=1}^n \sum_{j=1}^m \rho (X_i, Y_j) S_i^X S_j^Y - \frac{1}{\hat{n}^2} \sum_{i=1}^n \sum_{j=1}^n \rho (X_i, X_j) S_i^XS_j^X - \frac{1}{\hat{m}^2} \sum_{i=1}^n \sum_{j=1}^n \rho (Y_i, Y_j) S_i^Y S_j^Y.
\end{align}

The following theorem states that, as expected, the complete-case test statistic has the same asymptotic distribution as the complete-sample one, under the MCAR data. This result might seem obvious, but papers such as those by \cite{aleksic2023etAl} or \cite{aleksic2024impute} have shown us that the formal proofs can be very challenging.
\begin{theorem}\label{theorem_energy_cc}
    Let $X_1, X_2, \dots, X_n$ and $Y_1, Y_2, \dots, Y_m$ be two samples of random vectors from $\mathbb{R}^d$, such that the variables $X_1, \dots, X_n, Y_1, \dots, Y_m$ are independent and identically distributed with the characteristic function $\varphi (t)$. Let $T_{nm}$ and $T_{nm}^{CC}$ be as in \eqref{energy_test_statistic} and \eqref{T_CC}, respectively. Finally, assume that both samples have equal probabilities of a case being complete, i.e. $\mathbf{E}S^X = \mathbf{E}S^Y = q$. Then, under MCAR it holds that $\frac{nm}{n+m} T_{nm}$ and $\frac{\hat{n}\hat{m}}{\hat{n} + \hat{m}} T_{nm}^{CC}$ have the same asymptotic distribution. More precisely,
    \begin{align*}
        \frac{nm}{n+m} T_{nm} \overset{D}{\to} \| Z(t) \|^2_{w},  \quad \frac{\hat{n}\hat{m}}{\hat{n} + \hat{m}} T_{nm}^{CC} \overset{D}{\to} \| Z(t) \|^2_{w}, \quad \quad  \text{ as} \quad n, m \to \infty, \quad \frac{n}{n+m} \to \lambda^2,
    \end{align*}
    where $\{Z(t)\mid t \in \mathbb{R}^d \}$ is a centered Gaussian process with covariance operator defined for $s, t \in \mathbb{R}^d$ as 
    \begin{align}
        C(t,s) &= \mathbf{E} \left(  Z(s) Z (t) \right) \nonumber\\
        &=  \mathrm{Re} \left(\varphi (s-t)\right) + \mathrm{Im} \left(\varphi (s+t)\right)  \nonumber \\
        & \hspace{15pt}  - \mathrm{Re} \left(\varphi (t)\right) \mathrm{Re} \left(\varphi (s)\right)  - \mathrm{Im} \left(\varphi (s)\right)\mathrm{Re} \left(\varphi (t)\right)  - \mathrm{Im} \left(\varphi (s)\right) \mathrm{Im} \left(\varphi (t)\right) .\label{covariance_function}
    \end{align}
    Here, $\langle f, g \rangle_w = \int_{\mathbb{R}^d} f(t) g(t) w(t) \mathrm{d} t$, and $\| f \|_w^2 = \langle f, f \rangle_w$, where $w(t) = \| t\|^{d-1}$.
\end{theorem}
The proof can be found in the Appendix \ref{appendix_cc}.

In the same way as the original (complete-sample) energy test, the asymptotic distribution of the test statistic under the null hypothesis depends on the underlying distribution of the data and is therefore not distribution-free, so the bootstrap algorithm is utilized for calculating the critical or p-values of the test. The two proposed resampling procedures will be presented in the Subsection \ref{subsec:resampling}.

Due to its simplicity and low resource consumption, complete-case analysis became one of the most commonly used approaches when conducting various statistical analyses on incomplete datasets. Generally speaking, it can be a quick fix when data are MCAR, or the missingness rate is very low. Furthermore, for some procedures, such as independence testing \citep{aleksic2023etAl}, or multivariate normality testing 
\citep{aleksic2024impute}, complete-case analysis performed better under MCAR than some imputation methods. However, when either the data are not MCAR, or the missingness rate is very high, complete-case is known to leave much to be desired, producing biased estimates, decreasing the power of the test, having no type I error control, and much more \citep[e.g.][and others]{aleksic2024impute, tsatsi2024multivariate}. This is especially notable when dealing with high-dimensional data, where restricting attention to complete cases can, obviously, result in a significant loss of information and efficiency. The energy test is not an exception to this rule. For example, some simulations done by \cite{zeng2024mmd} indicate that, under their specific \textit{MMD-Miss} approach and MNAR setting, complete-case analysis has type I error asymptotically equal to 1.

Having the aforementioned flaws of complete-case analysis, wastefulness and sensitivity to data not being MCAR, it is imperative to seek for better approaches for conducting the energy test on incomplete samples, which could, at least partially, overcome them. 
As our simulations will demonstrate, complete case analysis can, in many cases, including the MCAR setting, be outperformed by certain weighting and imputation methods when appropriate bootstrap resampling is employed.

\subsection{Weighting methods}

As we have discussed, partially observed cases, although incomplete, could still carry useful information about the underlying structure of the data, or parameters of interest, and should not be disregarded outright. A natural way to incorporate these cases is to assign them weights based on the amount of observed information they contain. This approach ensures that observations with more observed components have a proportionally greater impact on the test statistic, hence making fuller use of the available data while acknowledging varying degrees of completeness across cases. For that purpose, we modify the original (de facto Euclidean) distance into a \emph{weighted distance} as
\begin{align}
    \rho_W \big((X, R^X), (Y, R^Y)\big) = \rho_{\mathrm{trunc}} \big((X, R^X), (Y, R^Y)\big) \cdot \frac{\sum_{i=1}^d (R^X \odot R^Y)_i}{d},
\end{align}
where $(R^X \odot R^Y)_i$ is the $i$th component of $R^X \odot R^Y$, and distance $\rho_{\mathrm{trunc}} \big((X, R^X), (Y, R^Y)\big)$ is calculated between those subvectors of $X$ and $Y$ where both are observed. The weight $\frac{1}{d} \sum_{i=1}^d (R^X \odot R^Y)_i$ is assigned so that the distance between the complete cases has weight 1, and the more missingness there is, the smaller the weight, and it contributes less to the overall sum. Naturally, \emph{weighted test statistic} is defined as
\begin{align}\label{T_W}
    T_{nm}^{W} =  \frac{2}{nm} \sum_{i=1}^n \sum_{j=1}^m \rho_W \big((X_i, R^X_i), &(Y_j, R^Y_j)\big)   - \frac{1}{n^2} \sum_{i=1}^n \sum_{j=1}^n \rho_W \big((X_i, R^X_i), (X_j, R^X_j)\big) \nonumber \\  
    &- \frac{1}{m^2} \sum_{i=1}^n \sum_{j=1}^n \rho_W \big( (Y_i, R_i^Y), (Y_j, R_j^Y) \big) .
\end{align}

Unlike the statistics $T_{nm}$ and $T_{nm}^{CC}$, there is no representation analogous to \eqref{chen_representation} for  the statistic $T_{nm}^{W}$. 
An alternative approach would be to express it as

\begin{align}
    T_{nm}^{W}  &= \frac{1}{n^2m^2} \sum_{i = 1}^n \sum_{j=1}^n \sum_{k=1}^m \sum_{l=1}^m \bigg(   \rho_W \big( (X_{i}, R_{i}^X), (Y_{l}, R_{l}^Y) \big)  \nonumber 
    + \rho_W \big( (X_{j}, R_{j}^X), (Y_{k}, R_{k}^Y) \big)    \nonumber \\
    &\hspace{125pt} - \rho_W \big( (X_{i}, R_{i}^X), (X_{j}, R_{j}^X) \big) - \rho_W \big( (Y_{k}, R_{k}^Y), (Y_{l}, R_{l}^Y) \big)  \bigg)  \nonumber \\
    &=: \frac{1}{n^2m^2} \sum_{i = 1}^n \sum_{j=1}^n \sum_{k=1}^m \sum_{l=1}^m h_W \big( (X_i, R_i^X), (X_j, R_j^X); (Y_k, R_k^Y), (Y_l, R_l^Y)   \big), \label{T_W_joined}
\end{align}
where 
\begin{align*}
    h_W \big( (X_1, R_1^X), (X_2, R_2^X); (Y_1, R_1^Y), (Y_2, R_2^Y)  \big) &= \rho_W \big( (X_{1}, R_{1}^X), (Y_{2}, R_{2}^Y) \big)  + \rho_W \big( (X_{2}, R_{2}^X), (Y_{1}, R_{1}^Y)\big) \nonumber \\
    &\hspace{25pt} - \rho_W \big( (X_{1}, R_{1}^X), (X_{2}, R_{2}^X) \big) - \rho_W \big( (Y_{1}, R_{1}^Y), (Y_{2}, R_{2}^Y) \big).
\end{align*}

$T_{nm}^{W}$ is itself a $V$-statistic, so we may proceed further in a standard manner by deriving the asymptotic distribution. Notice that, as an alternative, one can also consider the corresponding $U$-statistic. We state our conclusions as the following theorem, whose proof can be found in the Appendix \ref{appendix_weighted}.

\begin{theorem}\label{theorem_weighted}
    Let $X_1, \dots, X_n$ and $Y_1, \dots, Y_m$ be two independent samples of IID $d$-variate random vectors, and let $T_{nm}^W$ be as in \eqref{T_W_joined}. Let response indicators $R^X$ and $R^Y$ and their realizations $r^X$ and $r^Y$ be defined as in Section \ref{energy:novel_procedures}. 
    Define the integral operator $A : L^2 \big(\mathbb{R}^{2d}, \mathrm{d}F'(x,r^X) \big) \to L^2 \big(\mathbb{R}^{2d}, \mathrm{d}F'(x, r^X) \big)$, where $F'$ is a CDF of $(X, R^X)$ and of $(Y, R^Y)$, as
\begin{align}\label{A_operator_weighted}
    Ag\big( (x_1,r_1^X), (y_1, r_1^Y) \big) = \int_{\mathbb{R}^{2d}\times \mathbb{R}^{2d}} h_W \big( (x_1, r_1^X), (x_2, r_2^X); (y_1, r_1^Y), (y_2, r_2^Y)   \big) \mathrm{d} F'(x_2, r_2^X) \mathrm{d} F'(y_2, r_2^Y).
\end{align}
Let $\{\lambda_j, j \geq 1\}$ be the sequence of eigenvalues of $A$ and let $\{f_j, j \geq 1\}$ be the sequence of corresponding orthonormal eigenfunctions. Let
\begin{align*}
    c_j = \E \big[ f_j \big((X_1, R_1^X), (Y_1, R_1^Y) \big) f_j \big((X_1, R_1^X), (Y_2, R_2^Y) \big)\big],
\end{align*}
and
\begin{align*}
    \eta = \E \left[ \rho_W \left((X_1, R_1^X), (Y_1, R_1^Y)\right)  \right]. 
\end{align*}
    
If $(X_1,R_1^X) \overset{D}{=} (Y_1,R_1^Y)$, then
    \begin{align*}
        \frac{nm}{n+m} T_{nm}^W \overset{D}{\to} \eta\, + \, 2\sum_{j=1}^\infty \lambda_j c_j \left( \chi_{1,j}^2 - 1 \right) , \quad \text{as }\;\;n,m \to \infty, \quad \frac{n}{n+m} \to \lambda^2 \in (0,1),
    \end{align*}
where $\{\chi_{1,j}^2, j \geq 1\}$ are IID $\chi_1^2$-distributed random variables.
\end{theorem}

\begin{remark}
    Note that Theorem \ref{theorem_weighted} does not assume the MCAR assumption. However, it is clear that, if it holds, then it is sufficient to assume that $R^X \overset{D}{=} R^Y$ and that the null hypothesis holds.
\end{remark}

The null asymptotic distribution of the test statistic $T_{nm}^W$ clearly depends on the distribution of $(X_1, R_1^X)$. However, it is important to note that the use of random weights did not affect the type of the limiting distribution. 

Having the above said, turning to resampling procedures is a straightforward decision in this case as well. Those procedures will be presented in the Subsection \ref{subsec:resampling}.

\subsection{Imputation}

In many practical situations, a particular dataset will not be used exclusively for a single statistical procedure, such as two-sample testing, but rather as a subject of a broader range of analyses. This makes imputation particularly appealing: by filling in the missing values and producing a completed dataset, it allows analysts to apply standard methods, that are perhaps known not to be sensitive to ignoring the fact that the data are imputed, without the need to account for missingness at each step.  However, most of the standard methods are highly sensitive to that, and caution is needed when conducting such analysis. Moreover, in real-world scenarios, people that analyze the data may not have specialized knowledge of missing data techniques or access to tools that handle incomplete observations correctly. So the only option for them would be to treat the imputed dataset as complete.  For this reason, it is often desirable to provide a single, imputed version of the dataset that can be used in further analyses to follow, including hypothesis testing and many others. An algorithm that relies on imputing the data is a natural requirement for those types of scenarios. In a similar manner to the statistic $T_{nm}^W$,  we turn to the bootstrap once again. One such algorithm is proposed in  Subsection \ref{subsec:resampling}.

\subsection{Resampling procedures}\label{subsec:resampling}

We propose two bootstrap resampling procedures that can be used with both $T_{nm}^{CC}$ and $T_{nm}^W$, resulting in 4 different two-sample testing procedures overall. The first bootstrap approach, summarized in Algorithm \ref{algorithm_preserving} (replacing the generic $T$ with $T_{nm}^{CC}$ or $T_{nm}^W$), is designed to account for the structure of the incomplete data by treating complete and incomplete cases separately during resampling. Specifically, the pooled data are divided into two subsets: one containing only complete cases and the other containing only incomplete ones. Resampling is then carried out independently within each subset, after which the resampled complete and incomplete cases are recombined to form new bootstrap samples. This procedure ensures that the proportion of complete cases in each bootstrap sample remains close to the one in the original data. By preserving this proportion, the algorithm respects the original missingness pattern and avoids artificially inflating or deflating the amount of information available in the bootstrap replicates, compared to that in the original incomplete sample.

In addition to the first method, we also propose a simpler alternative, described in Algorithm \ref{algorithm_pooled}. Unlike the previous approach, this algorithm does not distinguish between complete and incomplete cases during resampling. Instead, it pools all available observations and randomly splits them into two bootstrap samples of fixed sizes $n$ and $m$. This method significantly reduces the complexity of the resampling procedure by avoiding the need to track and preserve the proportion of complete cases. It also slightly improves computation speed, which may not be crucial in real-world applications, but becomes very important when conducting simulation studies, where the testing procedure needs to be replicated tens or hundreds of thousands of times. 

However, this simplicity could come at a cost.  Since the proportion of complete to incomplete cases is not preserved across bootstrap samples, the testing procedures could have trouble controlling the type I error, or may exhibit reduced power in certain settings. One of the objectives of the simulation study that follows will be to examine whether this potential trade-off between computational simplicity and statistical performance has an impact in practice. In particular, we aim to assess whether the test remains well-calibrated and whether it retains sufficient power under various missingness scenarios.

\begin{algorithm}
  \caption{A bootstrap algorithm for the energy test: preserving the proportions of complete cases}\label{algorithm_preserving}
  \begin{algorithmic}[1]
    \State Start with incomplete samples ${x}=(x_1, \dots, x_{n})$ and ${y}=(y_1, \dots, y_{m})$ of $d$-variate vectors;
    \State Calculate the value $T({x}, {y})$ of the test statistic $T$;
    \State Produce two pooled samples: ${z}_{\mathrm{com}}$ that consists of complete cases from both ${x}$ and ${y}$, and ${z}_{\mathrm{inc}}$ that consists of incomplete cases;
    \State Randomly split ${z}_{\mathrm{com}}$ into ${x}^*_{\mathrm{com}}$ of size $\hat{n}$ and ${y}^*_{\mathrm{com}}$ of size $\hat{m}$; randomly split ${z}_{\mathrm{inc}}$ into ${x}^*_{\mathrm{inc}}$ of size $n - \hat{n}$ and ${y}^*_{\mathrm{inc}}$ of size $m - \hat{m}$; 
    \State Combine ${x}^*_{\mathrm{com}}$ and ${x}^*_{\mathrm{inc}}$ into ${x}^*$; combine ${y}^*_{\mathrm{com}}$ and ${y}^*_{\mathrm{inc}}$ into ${y}^*$; 
    \State Calculate $T^{*} = T ({x}^*, {y}^*)$;
    
    \State Repeat the steps 4-6 $B$ times to obtain $T_{1}^{*}, T_{2}^{*}, \dots, T_{B}^{*}$;
    \State Reject the null hypothesis for the significance level $\alpha$ if $T({x}, {y})$ is greater than the $(1 - \alpha)$-quantile of the empirical bootstrap distribution of $(T_{1}^{*}, T_{2}^{*}, \dots, T_{B}^{*})$.
  \end{algorithmic}
\end{algorithm}

\begin{algorithm}
  \caption{A bootstrap algorithm for the energy test: resampling directly from the pooled sample}\label{algorithm_pooled}
  \begin{algorithmic}[1]
    \State Start with incomplete samples ${x}=(x_1, \dots, x_{n})$ and ${y}=(y_1, \dots, y_{m})$ of $d$-variate vectors;
    \State Calculate the value $T({x}, {y})$ of the test statistic;
    \State Combine incomplete samples ${x}$ and ${y}$ to obtain the pooled sample ${z} = ({x}, {y})$;
    \State Randomly split pooled sample ${z}$ into ${x}^*$ of size $n$, and ${y}^*$ of size $m$;
    \State Calculate the value $T^{*} = T ({x}^*, {y}^*)$;
    \State Repeat the steps 4-5 $B$ times to obtain $T_{1}^{*}, T_{2}^{*}, \dots, T_{B}^{*}$;
    \State Reject the null hypothesis for the significance level $\alpha$ if $T({x}, {y})$ is greater than the $(1 - \alpha)$-quantile of the empirical bootstrap distribution of $(T_{1}^{*}, T_{2}^{*}, \dots, T_{B}^{*})$.
  \end{algorithmic}
\end{algorithm}

Under the imputation approach, we introduce the Algorithm \ref{algorithm_imputation}. The algorithm begins by imputing the original incomplete samples, resulting in their fully observed versions, from which the test statistic is computed. To approximate its null distribution via bootstrap, the algorithm pools the original data and resamples from it to create new incomplete bootstrap samples, which are then imputed using the same method. The test statistic is calculated on each imputed bootstrap pair, and the distribution of these replicates is used to determine the critical value. This approach allows the test to operate on completed data while remaining coherent with the imputation model throughout the resampling process. 

The success of this approach, however, depends critically on the quality of the imputation: poor or biased imputations may distort the type I error or reduce the power, as we will see from our simulations. Therefore, one of the goals of the simulation study will be to examine the sensitivity of this method to the choice of imputation strategy and to compare its performance with the other two algorithms. This will help clarify the trade-offs involved in choosing a more general-purpose, imputation approach over more specific, tailored, weighting methods.

\begin{algorithm}
  \caption{A bootstrap algorithm for the energy test: imputing the data}\label{algorithm_imputation}
  \begin{algorithmic}[1]
    \State Start with incomplete samples ${x}=(x_1, \dots, x_{n})$ and ${y}=(y_1, \dots, y_{m})$ of $d$-variate vectors;
    \State Impute the samples using the chosen method to obtain ${x}_{imp}$ and ${y}_{imp}$;
    \State Calculate the value $T_{nm}^{IMP}({x}_{imp}, {y}_{imp})$ of the test statistic $T_{nm}$ from \eqref{energy_test_statistic};
    \State Combine incomplete samples ${x}$ and ${y}$ to obtain the pooled sample ${z} = ({x}, {y})$;
    \State Randomly split pooled sample ${z}$ into ${x}^*$ of size $n$, and ${y}^*$ of size $m$;
    \State Impute ${x}^*$ and ${y}^*$ using the chosen method to obtain ${x}^*_{imp}$ and ${y}^*_{imp}$;
    \State Calculate the value $T_{nm}^{IMP, *} = T_{nm} ({x}^*_{imp}, {y}^*_{imp})$;
    \State Repeat the steps 5-7 $B$ times to obtain $T_{nm, 1}^{IMP, *}, T_{nm, 2}^{IMP, *}, \dots, T_{nm, B}^{IMP, *}$;
    \State Reject the null hypothesis for the significance level $\alpha$ if $T_{nm}^{IMP}({x}_{imp}, {y}_{imp})$ is greater than the $(1 - \alpha)$-quantile of the empirical bootstrap distribution of $(T_{nm, 1}^{IMP, *}, T_{nm, 2}^{IMP, *}, \dots, T_{nm, B}^{IMP, *})$.
  \end{algorithmic}
\end{algorithm}

\section{Power study on synthetic data}\label{energy:power_study}

In this section, we present the results of an empirical study conducted to assess the performance of the proposed methods. 

\subsection{Design of the study}\label{subsec:energy_study_design}

Due to the high computational demands of these methods, a trade-off had to be made by using the warp-speed modifications \citep{giacomini2013warp} of Algorithms \ref{algorithm_preserving}, \ref{algorithm_pooled}, and \ref{algorithm_imputation}, with $N = 5000$ Monte Carlo replicates.  The sample sizes were fixed to be $n = 100$ and $m=50$. Trivariate and decavariate data were considered to capture performance across different data dimensions, within the constraints of available computational resources.

\subsubsection*{Missingness mechanisms}

Regarding the missingness mechanisms, one MCAR mechanism is considered, as well as three MAR mechanisms. The first two are \textit{MAR 1 to 9} and \textit{MAR rank} mechanisms, implemented in the \texttt{R} package \texttt{missMethods} \citep{missMethods}, and have been previously used in papers by \cite{bordino2024tests}, \cite{aleksic2024novel, aleksic2025generalization}, and others. An additional MAR mechanism, referred to as the \textit{MAR logistic} mechanism, is considered. This mechanism operates by partitioning the variables into control variables and those subject to missingness. Missingness probabilities are then computed using a logistic regression model, with the control variables serving as covariates. This mechanism, although natural, poses a challenge when it comes to tuning the missingness rate, as the average proportion of missing data inherently depends on the distribution of the data itself. To address this, we employed the following workaround: for several choices of regression parameters, missingness is induced on data drawn from the standard normal distribution. Parameters that yielded a satisfactory missingness rate are then selected and subsequently used to generate missingness for all other data distributions. Naturally, this implies that the resulting missingness probability may vary across different distributions. For clarity, average missingness rates are included in all of the tables that present the results for this mechanism.

As shown in Tables \ref{tab:energy:legend} and \ref{tab:energy:legend_logistic}, which serve as a legend for the other Tables, combining Algorithms \ref{algorithm_preserving} and \ref{algorithm_pooled} with both $T_{nm}^{CC}$ and $T_{nm}^W$ yields four testing procedures. In addition, Algorithm \ref{algorithm_imputation} is paired with commonly used imputation methods: mean and median imputation \citep[from \texttt{R} package \texttt{missMethods} by][]{missMethods}, $k$-nearest neighbors ($k$NN) imputation with the median of the nearest neighbors \cite[from \texttt{R} package \texttt{bnstruct} by][]{franzin2017bnstruct}, and missForest \citep[from \texttt{R} package \texttt{missForest} by][]{stekhoven2013package}. This leaves us with a total of eight testing procedures to be evaluated in this study.

\subsubsection*{Trivariate data}

We focus on two of the most commonly studied types of distributional differences: shifts in location and changes in variance. The matrices
\begin{align*} 
C_1=\begin{bmatrix}
    0.5 & 0 & 0 \\ 0 & 0.5 & 0 \\ 0 & 0 & 0.5
\end{bmatrix} \quad \text{and} \quad C_2 = \begin{bmatrix}
    1 & 0.5 & 0.5 \\ 0.5 & 1 & 0.5 \\ 0.5 & 0.5 & 1
\end{bmatrix} ,
\end{align*}
together with the standard identity matrix, are used as covariance matrices, to capture the departure between as many covariance patterns as possible. Besides the standard zero mean (labeled with $0$), we consider the mean vector $m_1 = (0.5, 0.5, 0.5)$ to see how the testing procedures detect the change in mean. To examine both light and heavy-tailed distributions, we consider 5 non-degenerate normal distributions, as well as Student's $t$ distribution with 5 degrees of freedom. To be more specific, the distributions are the following: $\mathcal{N}(0, I)$, $\mathcal{N}(0, C_1)$, $\mathcal{N}(m_1, C_1)$, $\mathcal{N}(0, C_2)$, $\mathcal{N}(m_1, C_2)$, and $t_5(0, I)$, where $I$ denotes the identity matrix.

For MCAR, MAR 1 to 9, and MAR rank mechanisms, the missingness probability for each variable is set to $0.1$ and $0.4$, in order to examine both moderate and high missingness rates. Under the MCAR mechanism, all three variables are subject to missingness. Under the MAR mechanisms, only the second and third variables are allowed to be incomplete, as the first variable needs to be fully observed in order to govern the missingness of the others. For MAR logistic mechanism, we use the two sets of parameters. First, the intercept is set to be equal to $-5$, and slope parameters are $-1.9$ and $-1.5$, respectively, which results in an overall average missingness rate of around $10\%$ per each incomplete variable, for the data drawn from the standard normal distribution. Additionally, the intercept equal to $-1.05$ is used in combination with the slope parameters equal to $-1.7$ and $-0.6$, resulting in an overall average missingness rate of around $37\%$ per each incomplete variable, for the data distributed according to the standard normal distribution. For the other distributions, the missingness rates in both cases are shown in Tables \ref{tab:energy:powers_MAR_logistic_3D_m6_m1p9_m1p5} and \ref{tab:energy:powers_MAR_logistic_3D_m1p05_m1p7_m0p6}.

\subsubsection*{Decavariate data}

Similarly to the trivariate case, we consider the following distributions: $\mathcal{N}(0, I)$, $\mathcal{N}(0, C_3)$, $\mathcal{N}(m_2, C_3)$, $\mathcal{N}(0, C_4)$, $\mathcal{N}(m_2, C_4)$, $t_5(0, I)$, and $t_9(0, I)$, where
\begin{align*}
C_3=\begin{bmatrix}
    0.5 & 0 & \cdots & 0 \\ 0 & 0.5 & \cdots & 0 \\ \vdots & \vdots & \ddots & \vdots  \\0 & 0 & \cdots  & 0.5
\end{bmatrix}_{10 \times 10} \quad \text{and} \quad C_4 = \begin{bmatrix}
    1 & 0.5 & \cdots & 0.5 \\ 1 & 0.5 & \cdots & 0.5 \\ \vdots & \vdots & \ddots & \vdots  \\0.5 & 0.5 & \cdots  & 1
\end{bmatrix}_{10 \times 10},
\end{align*}
and $m_2 = (0.5, 0.5, \dots, 0.5) \in \mathbb{R}^{10}$.

The same missingness mechanisms are used as in the trivariate case. For the MCAR, MAR 1 to 9, and MAR rank mechanisms, the missingness probability is set to 0.05 for each variable. Unlike in the trivariate setting, where the lowest missingness probability is 0.1, this lower value is chosen to account for the impact of dimensionality. Specifically, if $p$ is the per-variable missingness probability, and $d$ is the data dimension, the probability for a case to be complete is $(1-p)^d$. Therefore, to maintain a comparable proportion of complete cases across dimensions, a lower per-variable missingness rate is used for higher-dimensional data.

For the MAR logistic mechanism, the first five variables are used as covariates for the logistic regression model, which imposes missingness in the other five variables. The intercept is set to be equal to $-2.6$, and the slope parameters are $0.3$, $1.8$, $1.7$, -$1.2$, and $1.9$, respectively. This choice of parameters results in the average missingness probability of $0.25$ per each incomplete variable, for the standard normal distribution.

\subsection{Results}

With the structure of the study now clearly laid out, we turn to presenting and discussing the results of the simulations. Given the volume of output, the results are organized in a series of tables. To avoid overloading the main text, these tables are placed in Appendix \ref{appendix_results}, and we refer to them throughout this section when needed.

\subsubsection*{MCAR data}

Table \ref{tab:energy:powers_MCAR_01} shows empirical type I errors and powers for trivariate data under the MCAR mechanism, with the missingness probability being equal to $0.1$ for each variable. As one can see, all eight testing procedures are essentially well calibrated.  The same holds for the missingness probability of $0.4$, as seen from the Table \ref{tab:energy:powers_MCAR_04}. Moreover, type I error appears to be even better in that case. In the setting with unequal missingness probabilities ($0.1$, $0.2$, and $0.3$ across variables), shown in Table \ref{tab:energy:powers_MCAR_01_02_03}, some of the methods exhibit slightly inflated type I error, but remain acceptably calibrated overall.

Analyzing the empirical power, $k$NN imputation is not the first choice in this setting, having the empirical power substantially lower than any other. Procedures that use the weighted test statistic performed the best overall. Mean and median imputation follow closely in some settings, but are mostly lacking power compared to the former two. 
Imputation using the missForest algorithm generally performs best out of all considered imputation methods. 
\color{black}
As expected, and seen from Tables \ref{tab:energy:powers_MCAR_01} and \ref{tab:energy:powers_MCAR_04}, the Algorithms \ref{algorithm_preserving}  and \ref{algorithm_pooled} perform basically the same; the simpler algorithm is even slightly better.

For decavariate data, basically the same conclusions can be made, wit the exception of $k$NN imputation, which performs comparably to the others.

Having everything previously said in mind, it is recommended, under the MCAR data, to use Algorithm \ref{algorithm_pooled} combined with the weighted test statistic $T_{nm}^{W}$, to obtain the best overall performance in terms of empirical type I error and power.

\subsubsection*{MAR data}

Results for the trivariate data under the MAR 1 to 9 mechanism with a missingness rate of $0.1$ are shown in Table \ref{tab:energy:powers_MAR1to9_01}. As the results indicate, all testing procedures properly control the type I error. When it comes to power, the weighting methods clearly perform best.  However, when the missingness rate increases to $0.4$ (Table \ref{tab:energy:powers_MAR1to9_04}), the weighting methods begin to lose control over the type I error. The complete-case approach shows similar issues, although to a lesser extent. Imputation methods retain control over the type I error in this scenario. Since not all methods maintain valid error rates, a fair comparison of power is not possible. However, among the imputation approaches, mean and median imputation clearly perform best when the distributional differences are due to the shift in location. The  $k$NN imputation consistently underperforms and should be avoided. The missForest imputation yields substantially better results than $k$NN.

The trivariate data under the MAR rank mechanism (Tables \ref{tab:energy:powers_MARrank_01} and \ref{tab:energy:powers_MARrank_04}) have none of issues with weighting methods losing control over the type I error. The power comparisons between testing procedures remain consistent.

Tables \ref{tab:energy:powers_MAR_logistic_3D_m6_m1p9_m1p5} and \ref{tab:energy:powers_MAR_logistic_3D_m1p05_m1p7_m0p6} present the results for the trivariate data under the MAR logistic mechanism for two different parameter settings. As seen from the tables, all testing procedures correctly control the type I error. In most scenarios, the weighting-based procedures consistently outperform the others. There are a few exceptions; for instance, in the comparison of $\mathcal{N}(0, I)$ and $\mathcal{N}(m_1, C_2)$, where mean or median imputation methods yield comparable power to the weighting methods. However, these cases involve clear differences in means, which would typically be picked up even by simpler tests designed to detect mean shifts. Notably, this behavior appears primarily under high missingness rates and is not observed when the missingness is moderate or low. When comparing the remaining methods, the complete-case approach generally outperforms imputation across most settings, most notably in scenarios where the means are equal.    
 
In the decavariate case, all tests appear to be well calibrated across the MAR scenarios considered. Although the weighting methods still exhibit a slight inflation of the type I error under MAR 1 to 9, their calibration remains within acceptable bounds. The relative ordering of the testing procedures in terms of power remains consistent with earlier findings. Notably, for distributions that differ clearly in their means, the advantage of imputation-based approaches becomes significantly more pronounced.
For some MAR mechanisms, an increase in dimensionality makes the $k$NN imputation work much better compared to the trivariate case. However, as seen from Table \ref{tab:energy:10D_powers_MAR_logistic}, it may not always be the case.

\section{Power study on real data}
\label{power_study_real}

In order to gain insight into the behavior of the proposed testing procedures on real data, we consider the \texttt{wine\_quality} data sets \citep{wine_quality_186}. These data sets contain 12 features describing various chemical properties related to wine quality. However, in our study, we focus on three features that are expected to exhibit smaller distributional differences between red and white wines: \texttt{pH}, \texttt{sulphates}, and \texttt{alcohol}.

\begin{figure}
    \centering
    \includegraphics[width=\textwidth]{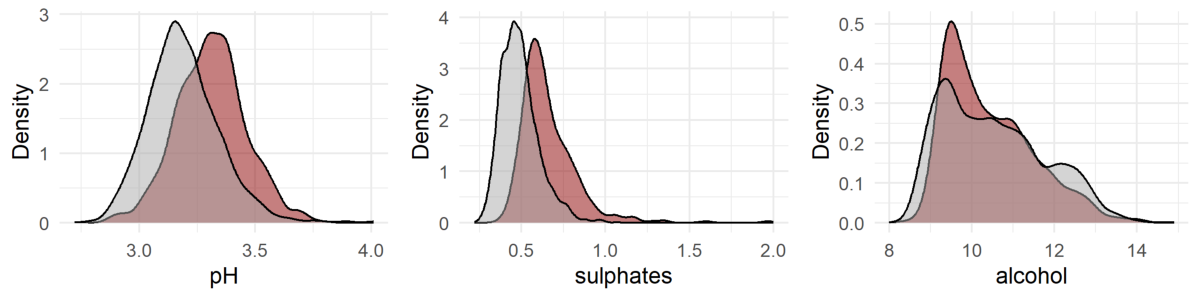}
    \caption{Marginal density plots for the entire populations of white (in gray) and red (in red) wines}
    \label{fig:densities}
\end{figure}

Figure \ref{fig:densities} presents marginal density plots for three selected variables across the entire population. As evident from the plots, the distributions of \texttt{pH} and \texttt{sulphates} clearly differ between the two wine types. While the distribution of \texttt{alcohol} appears similar across the groups, the overall joint distribution clearly varies between the wine types.

\subsection{Design of the study}

The analysis is conducted as follows: the \texttt{wine\_quality\_white} dataset (of size 4898) is treated as a population from which we draw IID samples of size $n = 100$, while the \texttt{wine\_quality\_red} dataset (of size 1599) is treated as a separate population from which we draw IID samples of size $m = 50$. We consider all missingness scenarios described in Section \ref{energy:power_study}. Specifically, MCAR missingness is generated with overall missingness probabilities of $0.1$ and $0.4$, as well as with unequal missingness probabilities set to $(0.1, 0.2, 0.3)$ across variables. Under the MAR 1 to 9 and MAR rank mechanisms, the first variable (\texttt{pH}) governs the missingness in \texttt{sulphates} and \texttt{alcohol}. As before, missingness probabilities of $0.1$ and $0.4$ are applied to each variable subject to missingness. 

Similarly to the synthetic data setting, where a standard normal distribution was used to tune regression parameters in order to achieve a specific missingness rate, the same approach is applied here by sampling both groups from the \texttt{wine\_quality\_white} population. Setting the intercept to $-1.3$ and the slope parameters to $0.7$ and $1.1$ yields a missingness rate of approximately $7\%$ for each incomplete variable in the white wine data. Similarly, setting the intercept and slope parameters to $9$, $-2$, and $-1$, respectively, results in a missingness rate of around $37\%$. The average missingness probabilities for samples drawn from different wine populations, as well as for those both drawn from the red wine population, are presented as underlined values in Table \ref{tab:energy:winequality_powers_MARlogistic}, in the same way as for the synthetic data.

The type I errors and powers are estimated as before, using Algorithms \ref{algorithm_preserving}, \ref{algorithm_pooled} and \ref{algorithm_imputation}, having the same 8 testing procedures to compare.

\subsection{Results}

Table \ref{tab:energy:winequality_powers_MCAR_01_04} presents the results for the MCAR mechanism. All methods are well calibrated; however, the weighting methods clearly achieve the highest power. This advantage becomes even more pronounced at higher missingness rates, where both the complete-case approach and imputation suffer substantial power loss, while the weighting methods remain unaffected. Nearest-neighbor imputation is affected the most by this. For smaller missingness rates, $k$NN remains the worst performer, but its performance is comparable to that of other imputation methods and the complete-case approach. 

Results for the MAR 1 to 9 mechanism are presented in Table \ref{tab:energy:winequality_powers_MAR1to9_01_04}. As in the synthetic data case, the weighting methods exhibit a slight inflation of the type I error. Nevertheless, they remain the preferred approach, as both the complete-case and imputation methods lose power when the missingness rate increases. This slight inflation of the type I error is not observed for the MAR rank mechanism, as shown in Table \ref{tab:energy:winequality_powers_MARrank_01_04}. In terms of power, the conclusions remain unchanged. For the MAR logistic mechanism (Table \ref{tab:energy:winequality_powers_MARlogistic}), all testing procedures are well calibrated. Once again, the weighting methods are the only approach that does not experience a significant loss of power as the missingness rate increases.

Unlike in the synthetic data case, the marginal distributions of \texttt{pH}, \texttt{sulphates}, and \texttt{alcohol} are clearly asymmetric, resulting in different performances for mean and median imputation. Across all missingness scenarios, mean imputation loses power more slowly than median imputation as the missingness rate increases. Moreover, mean imputation is the preferred method among all imputation approaches studied.

\section{Reflections and future work}\label{sec:reflections}

As shown, energy-based two-sample tests, when equipped with appropriate modifications, prove to be effective in the presence of missing data. Among all examined setups, the weighted approach demonstrated a particular advantage, as it allows the use of all available observations. This is especially valuable because the approach can be easily applied to other distance-based tests, which extends its practical relevance.
The favorable performance observed with increasing data dimension naturally raises the question of what happens as the dimension tends to infinity, an important open problem that remains to be explored in future research.
When it comes to the imputation approach, particular care should be taken in selecting the method, as some well-known techniques, such as $k$NN, may substantially reduce the test's power.

Last but not least, since two-sample tests serve as a foundation for nonparametric change-point detection \citep[see, e.g.][]{huvskova2006change, wei2022online, lukic2024change}, the presented ideas could be adapted to that context as well. To the best of our knowledge, such an extension has not yet been explored.

\section*{Disclosure statement}

The authors have no conflicts of interests to disclose.

\section*{Funding}

The work of D.G. Aleksić is supported by the Ministry of Science, Technological Development and Innovations of the Republic of Serbia (the contract 451-03-137/2025-03/200151). The work of B. Milošević is supported by the Ministry of Science, Technological Development and Innovations of the Republic of Serbia (the contract 451-03-136/2025-03/200104).

\section*{Notes on contributors}

D.G. Aleksić and B. Milošević have contributed equally to this work.


\appendix

\newpage
\section{Proof of Theorem \ref{theorem_energy_cc}}\label{appendix_cc}

We adapt the proof of Theorem 1 from \cite{chen2019some}, where they treated the problem of convergence of $T_{nn}$, i.e. the case of $m = n$. 

It is a known result \cite[see, e.g.][]{chen2019some} that 
\begin{align}\label{chen_representation}
    T_{mn} = \int_{\mathbb{R}^d} | \varphi_n (t) - \varphi_m (t)  |^2 w(t) \mathrm{d}t,
\end{align}
where $w(t) = \| t\|^{d-1}$ and $\varphi_n (t) = \frac{1}{n} \sum_{i=1}^n e^{it^T X_i}$ is an empirical characteristic function of a sample $X_1, X_2, \dots, X_m$ and $\varphi_m$ is defined similarly. 

Now, it is readily seen that
\begin{align}
    \frac{\hat{n}\hat{m}}{\hat{n}+\hat{m}}&T_{nm}^{CC} = \frac{\hat{n}\hat{m}}{\hat{n}+\hat{m}} \int_{\mathbb{R}^d} \left|  \frac{1}{\hat{n}} \sum_{i=1}^n e^{it^T X_i} S_i^X - \frac{1}{\hat{m}} \sum_{j=1}^m e^{it^T Y_j}S_j^Y \right|^2 \!\!\!w(t) \mathrm{d}t \nonumber \\
    &= \frac{\hat{n}\hat{m}}{\hat{n}+\hat{m}} \int_{\mathbb{R}^d} \left|  \left[ \frac{1}{\hat{n}} \sum_{i=1}^n \cos (t^T X_i) S_i^X - \frac{1}{\hat{m}} \sum_{j=1}^m \cos (t^TY_j)S_j^Y \right] \right. \nonumber \\ 
    & \hspace{150pt} \left. + i \left[ \frac{1}{\hat{n}} \sum_{i=1}^n \sin (t^T X_i)S_i^X - \frac{1}{\hat{m}} \sum_{j=1}^m \sin (t^T Y_j)S_j^Y \right] \right|^2\!\!\! w(t) \mathrm{d}t \nonumber \\
    &= \frac{\hat{n}\hat{m}}{\hat{n}+\hat{m}} \int_{\mathbb{R}^d} \left(  \left[ \frac{1}{\hat{n}} \sum_{i=1}^n \cos (t^T X_i) S_i^X - \frac{1}{\hat{m}} \sum_{j=1}^m \cos (t^TY_j)S_j^Y \right]^2 \right. \nonumber \\     
    & + \underbrace{2 \left[ \frac{1}{\hat{n}} \sum_{i=1}^n \cos (t^T X_i) S_i^X - \frac{1}{\hat{m}} \sum_{j=1}^m \cos (t^TY_j)S_j^Y \right]  \left[ \frac{1}{\hat{n}} \sum_{i=1}^n \sin (t^T X_i)S_i^X - \frac{1}{\hat{m}} \sum_{j=1}^m \sin (t^T Y_j)S_j^Y \right]}_{\text{Odd function of $t$, integrates to zero}} \nonumber \\
    &\left. \hspace{150pt} + \left[ \frac{1}{\hat{n}} \sum_{i=1}^n \sin (t^T X_i)S_i^X - \frac{1}{\hat{m}} \sum_{j=1}^m \sin (t^T Y_j)S_j^Y \right]^2 \right)\! w(t) \mathrm{d}t  \nonumber \\
    &= \frac{\hat{n}\hat{m}}{\hat{n}+\hat{m}} \int_{\mathbb{R}^d} \left[ \frac{1}{\hat{n}} \sum_{i=1}^n \cos (t^T X_i) S_i^X - \frac{1}{\hat{m}} \sum_{j=1}^m \cos (t^TY_j)S_j^Y \right. \nonumber \\
    & \hspace{150pt} \left.  + \frac{1}{\hat{n}} \sum_{i=1}^n \sin (t^T X_i)S_i^X - \frac{1}{\hat{m}} \sum_{j=1}^m \sin (t^T Y_j)S_j^Y \right]^2 \!\!\!w(t) \mathrm{d}t \nonumber \\
&= \frac{\hat{n}\hat{m}}{\hat{n}+\hat{m}} \int_{\mathbb{R}^d} \left[ \frac{1}{\hat{n}} \sum_{i=1}^n \left( \cos (t^T X_i) + \sin (t^T X_i) \right)S_i^X - \frac{1}{\hat{m}} \sum_{j=1}^m \left( \cos (t^T Y_j) + \sin (t^T Y_j) \right)S_j^Y  \right]^2\!\!\! w(t) \mathrm{d}t \nonumber \\
    &= \frac{\hat{n}\hat{m}}{\hat{n}+\hat{m}} \int_{\mathbb{R}^d} \bigg[ \frac{1}{\hat{n}} \sum_{i=1}^n \big[ \left( \cos (t^T X_i) + \sin (t^T X_i) - \left(\mathrm{Re} (\varphi (t)) + \mathrm{Im} (\varphi (t))  \right) \right)S_i^X \big] \nonumber \\
 & \hspace{200pt}+ \frac{n}{\hat{n}}\left(\mathrm{Re} (\varphi (t)) + \mathrm{Im} (\varphi (t))  \right)S_i^X  \nonumber \\
 &  \hspace{75pt}- \frac{1}{\hat{m}} \sum_{j=1}^m \big[  \left( \cos (t^T Y_j) + \sin (t^T Y_j) - \left(\mathrm{Re} (\varphi (t)) + \mathrm{Im} (\varphi (t))  \right) \right)S_j^Y \big] \nonumber \\
 & \hspace{200pt}- \frac{m}{\hat{m}}\left(\mathrm{Re} (\varphi (t)) + \mathrm{Im} (\varphi (t))  \right)S_j^Y     \bigg]^2 \!\!\!w(t) \mathrm{d}t \nonumber \\
 &= \frac{\hat{n}\hat{m}}{\hat{n}+\hat{m}} \int_{\mathbb{R}^d} \bigg[ \frac{1}{\hat{n}} \sum_{i=1}^n \big[ \left( \cos (t^T X_i) + \sin (t^T X_i) - \left(\mathrm{Re} (\varphi (t)) + \mathrm{Im} (\varphi (t))  \right) \right)S_i^X \big]  \nonumber \\
  &  \hspace{75pt}- \frac{1}{\hat{m}} \sum_{j=1}^m \big[  \left( \cos (t^T Y_j) + \sin (t^T Y_j) - \left(\mathrm{Re} (\varphi (t)) + \mathrm{Im} (\varphi (t))  \right) \right)S_j^Y \big]  \nonumber \\
 & \hspace{100pt}+ \frac{1}{\hat{n}} \sum_{i=1}^n\left(\mathrm{Re} (\varphi (t)) + \mathrm{Im} (\varphi (t))  \right)S_i^X - \frac{1}{\hat{m}}\sum_{j=1}^m\left(\mathrm{Re} (\varphi (t)) + \mathrm{Im} (\varphi (t))  \right)S_j^Y     \bigg]^2\!\!\! w(t) \mathrm{d}t \nonumber \\
 &=  \int_{\mathbb{R}^d} \bigg[ \sqrt{\frac{\hat{m}}{\hat{n}+\hat{m}}}\frac{1}{\sqrt{\hat{n}}} \sum_{i=1}^n \big[ \left( \cos (t^T X_i) + \sin (t^T X_i) - \left(\mathrm{Re} (\varphi (t)) + \mathrm{Im} (\varphi (t))  \right) \right)S_i^X \big]  \nonumber \\
  &  \hspace{75pt}- \sqrt{\frac{\hat{n}}{\hat{n}+\hat{m}}}\frac{1}{ \sqrt{\hat{m}}} \sum_{j=1}^m \big[  \left( \cos (t^T Y_j) + \sin (t^T Y_j) - \left(\mathrm{Re} (\varphi (t)) + \mathrm{Im} (\varphi (t))  \right) \right)S_j^Y \big]  \nonumber \\
 & \hspace{20pt}+\underbrace{\sqrt{\frac{\hat{n}\hat{m}}{\hat{n} + \hat{m}}} \left(\mathrm{Re} (\varphi (t)) + \mathrm{Im} (\varphi (t))  \right)  \frac{1}{\hat{n}}\underbrace{\sum_{i=1}^n S_i^X }_{= \hat{n}} - \sqrt{\frac{\hat{n}\hat{m}}{\hat{n}+\hat{m}}}\left(\mathrm{Re} (\varphi (t)) + \mathrm{Im} (\varphi (t))  \right)  \frac{1}{\hat{m}}\underbrace{\sum_{j=1}^m S_j^Y }_{= \hat{m}} }_{=0}   \bigg]^2 \!\!\!w(t) \mathrm{d}t \nonumber \\
 &=  \int_{\mathbb{R}^d} \left[ \sqrt{\frac{\hat{m}}{\hat{n}+\hat{m}}}\frac{1}{\sqrt{\hat{n}}} \sum_{i=1}^n \big[ \left( \cos (t^T X_i) + \sin (t^T X_i)\right) - \left(\mathrm{Re} (\varphi (t)) + \mathrm{Im} (\varphi (t))  \right)  \big] S_i^X \right. \nonumber \\
  &\left.  \hspace{75pt}- \sqrt{\frac{\hat{n}}{\hat{n}+\hat{m}}}\frac{1}{ \sqrt{\hat{m}}} \sum_{j=1}^m \big[  \left( \cos (t^T Y_j) + \sin (t^T Y_j)\right) - \left(\mathrm{Re} (\varphi (t)) + \mathrm{Im} (\varphi (t))  \right)  \big] S_j^Y \right]^2\!\!\! w(t) \mathrm{d}t\nonumber \\
  &=  \int_{\mathbb{R}^d} \left[ \frac{\sqrt{\frac{\hat{m}}{\hat{n}+\hat{m}}}\frac{1}{\sqrt{\hat{n}}}}{\sqrt{\frac{m}{n+m}} \frac{1}{\sqrt{n}}} \sqrt{q}\sqrt{\frac{m}{n+m}} \frac{1}{\sqrt{n}}\sum_{i=1}^n \left[ \left( \cos (t^T X_i) + \sin (t^T X_i) \right) - \left(\mathrm{Re} (\varphi (t)) + \mathrm{Im} (\varphi (t))  \right)  \right] \frac{S_i^X}{\sqrt{q}}\right. \nonumber \\
  & \left. \hspace{5pt}- \frac{ \sqrt{\frac{\hat{n}}{\hat{n}+\hat{m}}}\frac{1}{ \sqrt{\hat{m}}}}{\sqrt{\frac{n}{n+m}}\frac{1}{\sqrt{m}}} \sqrt{q} \sqrt{\frac{n}{n+m}}\frac{1}{\sqrt{m}} \sum_{j=1}^m \left[  \left( \cos (t^T Y_j) + \sin (t^T Y_j) \right) - \left(\mathrm{Re} (\varphi (t)) + \mathrm{Im} (\varphi (t))  \right) \right]\frac{S_j^Y}{\sqrt{q}} \right]^2\!\!\! w(t) \mathrm{d}t  \label{end_of_long_calc_energy_cc}
\end{align}

Now, by the Law of Large Numbers and the Continuous Mapping Theorem we have that
\begin{align}\label{q_constants}
    \frac{\sqrt{\frac{\hat{m}}{\hat{n}+\hat{m}}}\frac{1}{\sqrt{\hat{n}}}}{\sqrt{\frac{m}{n+m}} \frac{1}{\sqrt{n}}} \sqrt{q} \overset{P}{\to} 1, \quad \text{ and } \quad \frac{ \sqrt{\frac{\hat{n}}{\hat{n}+\hat{m}}}\frac{1}{ \sqrt{\hat{m}}}}{\sqrt{\frac{n}{n+m}}\frac{1}{\sqrt{m}}} \sqrt{q} \overset{P}{\to} 1.
\end{align}
By Slutsky's theorem, these terms can be asymptotically treated as 1.

Denote 
\begin{align*}
   Z_{n, 1}(t) := \frac{1}{\sqrt{n}}\sum_{i=1}^n \left[ \left( \cos (t^T X_i) + \sin (t^T X_i) \right) - \left(\mathrm{Re} (\varphi (t)) + \mathrm{Im} (\varphi (t))  \right)  \right] \frac{S_i^X}{\sqrt{q}}
\end{align*}
and
\begin{align*}
    Z_{m, 2}(t) := \frac{1}{\sqrt{m}} \sum_{j=1}^m \left[  \left( \cos (t^T Y_j) + \sin (t^T Y_j) \right) - \left(\mathrm{Re} (\varphi (t)) + \mathrm{Im} (\varphi (t))  \right) \right]\frac{S_j^Y}{\sqrt{q}}
\end{align*}
One can easily note that for every $1 \leq i \leq n$ and $1 \leq j \leq n$ random functions 
\begin{align*}
     h((X_i, R_i^X), t) :=  \left[ \left( \cos (t^T X_i) + \sin (t^T X_i) \right) - \left(\mathrm{Re} (\varphi (t)) + \mathrm{Im} (\varphi (t))  \right)  \right] \frac{S_i^X}{\sqrt{q}}
\end{align*} 
and 
\begin{align*}
     h((Y_j, R_j^Y), t) := \left[  \left( \cos (t^T Y_j) + \sin (t^T Y_j) \right) - \left(\mathrm{Re} (\varphi (t)) + \mathrm{Im} (\varphi (t))  \right) \right]\frac{S_j^Y}{\sqrt{q}}
\end{align*} 
are independent and identically distributed centered random elements of $L^2 (\mathbb{R}^d, w(t) \mathrm{d}t)$ with the  covariance function equal to 
\begin{align}
    \mathbf{E} & \bigg\{ \left[\left( \cos (t^T X_1) + \sin (t^T X_1) \right) - \left(\mathrm{Re} (\varphi (t)) + \mathrm{Im} (\varphi (t))  \right)  \right] \nonumber \\
 &\hspace{50pt}\cdot \left[\left( \cos (s^T X_1) + \sin (s^T X_1) \right) - \left(\mathrm{Re} (\varphi (s)) + \mathrm{Im} (\varphi (s))  \right)  \right] \frac{(S_1^X)^2}{q} \bigg\}\nonumber \\
 &=  \cdot \mathbf{E} \left[ \cos (t^T X_1) \cos (s^T X_1) + \cos (t^T X_1) \sin (s^T X_1)  + \sin (t^T X_1) \cos (s^T X_1) + \sin (t^T X_1) \sin (s^T X_1)\right]  \nonumber \\
 & \hspace{200pt} -  \left(\mathrm{Re} (\varphi (t)) + \mathrm{Im} (\varphi (t))  \right) \cdot  \left(\mathrm{Re} (\varphi (s)) + \mathrm{Im} (\varphi (s))  \right) \nonumber \\
 &=  \cdot \mathbf{E} \bigg[ \frac12 \bigg( \cos \left((t-s)^T X_1 \right) + \cos \left(((t+s)^T X_1 \right)\bigg) + \frac12 \bigg( \sin \left((s-t)^T X_1\right) + \sin \left((s+t)^T X_1\right)  \bigg) \nonumber \\
 &\hspace{30pt}+ \frac12 \bigg( \sin \left((t-s)^T X_1\right) + \sin \left((s+t)^T X_1\right)  \bigg)  + \frac12 \bigg( \cos \left((t-s)^T X_1\right) - \cos \left((t+s)^T X_1\right)  \bigg)\bigg] \nonumber \\
 & \hspace{200pt} -  \left(\mathrm{Re} (\varphi (t)) + \mathrm{Im} (\varphi (t))  \right) \cdot  \left(\mathrm{Re} (\varphi (s)) + \mathrm{Im} (\varphi (s))  \right) \nonumber \\
 &=  \mathrm{Re} (\varphi (t-s)) +  \mathrm{Im} (\varphi (t+s)) -\mathrm{Re} (\varphi (t)) \mathrm{Re} (\varphi (s)) -  \mathrm{Im} (\varphi (t)) \mathrm{Re} (\varphi (s)) \nonumber  \\
 & \hspace{250pt}-  \mathrm{Re} (\varphi (t)) \mathrm{Im} (\varphi (s)) -  \mathrm{Im} (\varphi (t)) \mathrm{Im} (\varphi (s)) ,\label{cov_fun_interm}
\end{align}
which is exactly the covariance function $C(t,s)$ from \eqref{covariance_function}. 

Now, since random functions $h((X_i, R_i^X), t)$ and $h((Y_j, R_j^Y), t)$ are elements of $L^2 (\mathbb{R}^d, w(t)\mathrm{d}t)$ with existing covariance function \eqref{cov_fun_interm}, we can apply the Central limit theorem for Hilbert spaces \citep[e.g.][Theorem 17.29]{henze2024asymptotic} to conclude that
\begin{align*}
   \sqrt{\frac{m}{n+m}} Z_{n,1}(t) \overset{D}{\to} Z_1 (t) \quad \text{ and } \quad \sqrt{\frac{n}{n+m}}Z_{m,2}(t) \overset{D}{\to} Z_2 (t),
\end{align*}
where $\{Z_1 (t) \mid t \in \mathbb{R}^d\}$ and $\{ Z_2(t) \mid t \in \mathbb{R}^d  \}$ are independent, centered Gaussian processes with covariance functions equal to $\lambda^2C(t,s)$ and $(1 - \lambda^2) C(t,s)$, respectively, where $C(t,s)$ is defined in \eqref{covariance_function}. Having the independence of $Z_1$ and $Z_2$ and the convergence \eqref{q_constants} we can conclude that the the difference inside the large square brackets in \eqref{end_of_long_calc_energy_cc} converges in distribution to the random process
\begin{align*}
     Z(t) = Z_1(t) -  Z_2 (t),
\end{align*}
with  the covariance function equal to the sum of corresponding covariance functions, due to independence:
\begin{align*}
    \lambda^2 C(t,s) + (1 - \lambda^2) C(t,s) = C(t,s).
\end{align*}
Recalling the definition (and continuity) of $\|\cdot\|_w$, we finally conclude that 
\begin{align*}
    \frac{\hat{n}\hat{m}}{\hat{n}+\hat{m}} T_{nm}^{CC} \overset{D}{\to} \| Z(t)\|^2_w,
\end{align*}
where $\{Z(t)\mid t\in \mathbb{R}^d\}$ is the centered Gaussian process with the covariance function $C(t,s)$ from \eqref{covariance_function}, which concludes this part of the proof.

The proof that $nm T_{nm}/(n+m)$ also converges to $\| Z(t)\|^2_w$ follows trivially from the fact that, when the data are complete, $\hat{n} = n, \hat{m}=m$ and $q = 1$, and is known from the literature \citep{chen2019some}. This concludes the proof of Theorem \ref{theorem_energy_cc}.

\newpage
\section{Results of Neuhaus (1977)}\label{appendix_neuhaus}

For the reader’s convenience, we restate here the results of \cite{neuhaus1977functional} that are relevant to our work. We note that only a subset of those results is presented, and that the original statements are slightly more general than given here. Furthermore, the notation has been substantially modified to suit our purposes.

Let us have two independent samples $X_1, X_2, \dots, X_n$ and $Y_1, Y_2, \dots Y_m$ of IID $d$-variate random vectors with the common CDF $F$, and let $h(x_1, x_2; y_1,y_2)$ be a measurable kernel symmetric with respect to the mutual permutations of $x_1$ and $x_2$, or $y_1$ and $y_2$, and let $\E \big( h(X_1, X_2; Y_1, Y_2)^2 \big) < \infty$. Additionally, assume that $h$ is a degenerate kernel in a way that
\begin{align*}
    h_{1,1} (x,y) = \E \bigg( h(X_1, X_2; Y_1,Y_2) \big| X_1 = x, \, Y_1 = y \bigg) = 0
\end{align*}
for almost every pair $(x,y)$, according to the distribution of $(X_1, Y_1)$. 

According to the Spectral theorem \cite[for a modern reference see, e.g.][Th. 8.14]{henze2024asymptotic}, there is an orthonormal sequence $f_1, f_2, \dots L^2 \big(\mathbb{R}^{2d}, \mathrm{d}F(x)\mathrm{d}F(y) \big)$, such that $\int f_j (x,y) \mathrm{d}F(x) \mathrm{d}F(y) = 0$ for $j \geq 1$, and a diminishing sequence $\lambda_1, \lambda_2, \dots $ of positive real numbers such that $\sum_{j=1}^\infty \lambda_j^2= \E \big( h(X_1, X_2; Y_1, Y_2)^2 \big) < \infty$, so that, if 
\begin{align*}
    h^s(x_1, x_2; y_1, y_2) = \sum_{j=1}^s \lambda_j f_j(x_1, y_1) f_j (x_2, y_2),
\end{align*}
then
\begin{align*}
    \lim_{s \to \infty} \E\left[ \left( h(X_1, X_2; Y_1, Y_2) - h^s(X_1, X_2; Y_1, Y_2) \right)^2 \right] = 0.
\end{align*}
Moreover, $\lambda_1, \lambda_2, \dots$ are eigenvalues of the integral operator $A : L^2 \big(\mathbb{R}^{2d}, \mathrm{d}F(x)\mathrm{d}F(y) \big) \to L^2 \big(\mathbb{R}^{2d}, \mathrm{d}F(x)\mathrm{d}F(y) \big)$ defined as
\begin{align*}
    Ag(x_1, y_1) = \int_{\mathbb{R}^{2d}} k(x_1, x_2; y_1, y_2) g(x_2, y_2) \mathrm{d} F(x_2) \mathrm{d} F(y_2),
\end{align*}
with orthonormal eigenfunctions $f_1, f_2, \dots$ and $k(x_1, x_2; y_1, y_2) = h(x_1, x_2; y_1, y_2) - \E [ h(X_1, X_2; Y_1, Y_2)]$.

Let $W_{1j}, W_{2j}$, $j \geq 1$, be independent Brownian motions on $[0,1]$. For any $0 \leq t_1, t_2 \leq 1$, let
\begin{align}\label{U_nm_neuhaus}
    U_{nm} (t_1, t_2) = \frac{1}{nm (n+m)} \sum_{i=1}^{\lceil nt_1 \rceil} \sum_{\substack{j = 1 \\ j \neq i}}^{\lceil nt_1 \rceil} \sum_{k=1}^{\lceil mt_2 \rceil} \sum_{\substack{l = 1 \\ l \neq k}}^{\lceil mt_2 \rceil}  h\big( X_{i}, X_{j}; Y_{k}, Y_{l} \big).
\end{align}
Let
\begin{align}\label{neuhaus_limiting_U}
    U(t_1, t_2) = \sum_{j=1}^\infty \lambda_j \bigg[ \big(a_j t_2 W_{1j}(t_1) + b_j t_1 W_{2j}(t_2)\big)^2  - \big( a_j^2t_2+b_j^2t_1 \big)t_1t_2 \bigg],
\end{align}
where 
\begin{align}\label{aj_bj_neuhaus}
    a_j^2 = \beta \int \left( \E f_j (x, Y_1)  \right)^2 \mathrm{d} F(x), \quad b_j^2 = \alpha \int \left( \E f_j (X_1, y)\right)^2 \mathrm{d} F(x), \quad j \geq 1,
\end{align}
and $\alpha$ and $\beta$ will be defined soon. We observe $U_{nm}$ as a random element in the space $D([0,1]^2)$ of all real functions on $[0,1]^2$ with no discontinuities of the second kind, equipped with the Skorohod metric \cite[see e.g.][Sec. 14.2]{billingsley1968convergence}. 

The following result is due to \cite{neuhaus1977functional}.

\begin{theorem}[Neuhaus, 1977]
As $n, m \to \infty$, $n/(n+m) \to \alpha$, $m/(n+m) \to \beta$, it holds that
\begin{align}\label{neuhaus_convergence}
    U_{nm} \overset{D}{\to}  U, \quad \text{in }D_2.
\end{align}.
\end{theorem}

\newpage
\section{Proof of Theorem \ref{theorem_weighted}}\label{appendix_weighted}

We first note that the assumptions of Theorem \ref{theorem_weighted} are consistent, since the operator $A$ from \eqref{A_operator_weighted} is indeed known to be compact and self-adjoint, so its eigenvalues do form a diminishing sequence, and the eigenfunctions are orthonormal \citep[see e.g.][Ch. 8]{henze2024asymptotic}.

The kernel is weakly degenerate. Indeed, it holds that
\begin{align*}
    h_{1,1} \big( (x_1, r_1^X), (y_1, r_1^Y) \big) &= \E \bigg[  \rho_W \big( (x_{1}, r_{1}^X), (Y_{2}, R_{2}^Y) \big)  + \rho_W \big( (X_{2}, R_{2}^X), (y_{1}, r_{1}^Y) \big) \nonumber \\
    &\hspace{25pt} - \rho_W \big( (x_{1}, r_{1}^X), (X_{2}, R_{2}^X) \big) - \rho_W \big( (y_{1}, y_{1}^Y), (Y_{2}, R_{2}^Y) \big) \bigg] \\
    &= 0,
\end{align*}
where two pairs of terms cancel out due to the symmetry of $\rho_W$ and the fact that $(X, R^X) \overset{D}{=} (Y, R^Y)$.

The key idea of the proof is to use the results of \cite{neuhaus1977functional} that we have restated in our notation in the Appendix \ref{appendix_neuhaus}. Similarly to the Lemma 2 of \cite{fernandez2008test}, it holds that, for every $j \geq 1$, $f_j\big((x, r^X), (y, r^Y)  \big) = - f_j \big( (y, r^Y), (x, r^X) \big)$, and, as a consequence,
\begin{align*}
    \E \big[ f_j \big((X_1, R_1^X), (Y_1, R_1^Y) \big) f_j \big((X_1, R_1^X), (Y_2, R_2^Y) \big)\big] = \E \big[ f_j \big((X_1, R_1^X), (Y_1, R_1^Y) \big) f_j \big((X_2, R_2^X), (Y_1, R_1^Y) \big)\big].
\end{align*}
Denote this quantity as $c_j$. If $a_j^2$ and $b_j^2$ are defined as in \eqref{aj_bj_neuhaus}, it is readily seen that
\begin{align*}
    a_j^2 &= (1 - \lambda^2) \E \big[ f_j \big((X_1, R_1^X), (Y_1, R_1^Y) \big) f_j \big((X_1, R_1^X), (Y_2, R_2^Y) \big)\big] = (1 - \lambda^2) c_j, \\ 
    b_j &= \lambda^2 \E \big[ f_j \big((X_1, R_1^X), (Y_1, R_1^Y) \big) f_j \big((X_2, R_2^X), (Y_1, R_1^Y) \big)\big] = \lambda^2 c_j,
\end{align*}
for every $j \geq 1$. 

If  $\frac{nm}{n+m}T_{nm}^W$ is understood in the context of \eqref{U_nm_neuhaus}, for $t_1 = t_2 = 1$, then the corresponding $U(1,1)$ can be written as
\begin{align*}
    U(1,1) &= \sum_{j=1}^\infty \lambda_j \bigg[ \big(a_j W_{1j}(1) + b_j W_{2j}(1)\big)^2  - \big( a_j^2+b_j^2 \big)\bigg] \\
    &= \sum_{j=1}^\infty \lambda_j c_j \bigg[ \big( \underbrace{W_{1j} (1) + W_{2j} (1)}_{\sim \mathcal{N}(0,2)} \big)^2 - 2   \bigg] \\
    &= 2\sum_{j=1}^\infty  \lambda_j c_j \left( \chi_{1,j}^2 - 1 \right).
\end{align*}

Next, observe that 
\begin{align*}
    \frac{nm}{n+m} T_{nm}^W &= \frac{1}{nm (n+m)} \sum_{i = 1}^n \sum_{j=1}^n \sum_{k=1}^m \sum_{l=1}^m h_W \big( (X_i, R_i^X), (X_j, R_j^X); (Y_k, R_k^Y), (Y_l, R_l^Y)   \big) \nonumber \\
    &= \frac{1}{nm (n+m)} \sum_{i = 1}^n \sum_{\substack{j=1\\ j \neq i}}^n \sum_{k=1}^m \sum_{\substack{l = 1}}^m h_W \big( (X_i, R_i^X), (X_j, R_j^X); (Y_k, R_k^Y), (Y_l, R_l^Y)   \big) \nonumber \\
    & \hspace{80pt}+  \frac{1}{nm (n+m)} \sum_{i = 1}^n  \sum_{k=1}^m \sum_{\substack{l = 1}}^m h_W \big( (X_i, R_i^X), (X_i, R_i^X); (Y_k, R_k^Y), (Y_l, R_l^Y)   \big)  \nonumber\\
    &= \frac{1}{nm (n+m)} \sum_{i = 1}^n \sum_{\substack{j=1\\ j \neq i}}^n \sum_{k=1}^m \sum_{\substack{l = 1 \\ l \neq k}}^m h_W \big( (X_i, R_i^X), (X_j, R_j^X); (Y_k, R_k^Y), (Y_l, R_l^Y)   \big) \nonumber \\
    & \hspace{80pt}+  \frac{1}{nm (n+m)} \sum_{i = 1}^n \sum_{\substack{j=1\\ j \neq i}}^n \sum_{k=1}^m  h_W \big( (X_i, R_i^X), (X_j, R_j^X); (Y_k, R_k^Y), (Y_k, R_k^Y)   \big)  \\
    & \hspace{80pt}+  \frac{1}{nm (n+m)} \sum_{i = 1}^n  \sum_{k=1}^m \sum_{\substack{l = 1}}^m h_W \big( (X_i, R_i^X), (X_i, R_i^X); (Y_k, R_k^Y), (Y_l, R_l^Y)   \big)  \nonumber\\
    &= \frac{1}{nm (n+m)} \sum_{i = 1}^n \sum_{\substack{j=1\\ j \neq i}}^n \sum_{k=1}^m \sum_{\substack{l = 1 \\ l \neq k}}^m h_W \big( (X_i, R_i^X), (X_j, R_j^X); (Y_k, R_k^Y), (Y_l, R_l^Y)   \big) \nonumber \\
    & \hspace{80pt}+  \frac{1}{nm (n+m)} \sum_{i = 1}^n \sum_{\substack{j=1\\ j \neq i}}^n \sum_{k=1}^m  h_W \big( (X_i, R_i^X), (X_j, R_j^X); (Y_k, R_k^Y), (Y_k, R_k^Y)   \big)  \\
    & \hspace{80pt}+  \frac{1}{nm (n+m)} \sum_{i = 1}^n  \sum_{k=1}^m \sum_{\substack{l = 1 \\ l \neq k}}^m h_W \big( (X_i, R_i^X), (X_i, R_i^X); (Y_k, R_k^Y), (Y_l, R_l^Y)   \big)  \nonumber\\
    & \hspace{80pt}+  \frac{1}{nm (n+m)} \sum_{i = 1}^n  \sum_{k=1}^m  h_W \big( (X_i, R_i^X), (X_i, R_i^X); (Y_k, R_k^Y), (Y_k, R_k^Y)   \big)  \nonumber\\
    &= \frac{1}{nm (n+m)} \sum_{i = 1}^n \sum_{\substack{j=1\\ j \neq i}}^n \sum_{k=1}^m \sum_{\substack{l = 1 \\ l \neq k}}^m h_W \big( (X_i, R_i^X), (X_j, R_j^X); (Y_k, R_k^Y), (Y_l, R_l^Y)   \big) \nonumber \\
    & \hspace{30pt}+  \left(\frac{n}{n+m} - \frac{1}{n+m}\right) \frac{1}{n(n-1)m} \sum_{i = 1}^n \sum_{\substack{j=1\\ j \neq i}}^n \sum_{k=1}^m  h_W \big( (X_i, R_i^X), (X_j, R_j^X); (Y_k, R_k^Y), (Y_k, R_k^Y)   \big) \\
    & \hspace{30pt}+  \left(\frac{m}{ n+m} - \frac{1}{n+m} \right) \frac{1}{nm(m-1)} \sum_{i = 1}^n  \sum_{k=1}^m \sum_{\substack{l = 1 \\ l \neq k}}^m h_W \big( (X_i, R_i^X), (X_i, R_i^X); (Y_k, R_k^Y), (Y_l, R_l^Y)   \big)  \nonumber\\
    & \hspace{30pt}+  \frac{1}{n+m} \frac{1}{nm} \sum_{i = 1}^n  \sum_{k=1}^m  h_W \big( (X_i, R_i^X), (X_i, R_i^X); (Y_k, R_k^Y), (Y_k, R_k^Y)   \big)  \nonumber\\
    &= M_{nm} + \left( \frac{n}{n+m} - \frac{1}{n+m} \right) N_{nm} + \left( \frac{m}{n+m} - \frac{1}{n+m} \right) P_{nm} + \frac{1}{n+m} Q_{nm},
\end{align*}
where
\begin{align*}
    M_{nm} = \frac{1}{nm (n+m)} \sum_{i = 1}^n \sum_{\substack{j=1\\ j \neq i}}^n \sum_{k=1}^m \sum_{\substack{l = 1 \\ l \neq k}}^m h_W \big( (X_i, R_i^X), (X_j, R_j^X); (Y_k, R_k^Y), (Y_l, R_l^Y)   \big),
\end{align*}
\begin{align*}
    N_{nm} = \frac{1}{n(n-1)m} \sum_{i = 1}^n \sum_{\substack{j=1\\ j \neq i}}^n \sum_{k=1}^m  h_W \big( (X_i, R_i^X), (X_j, R_j^X); (Y_k, R_k^Y), (Y_k, R_k^Y)   \big),
\end{align*}
\begin{align*}
    P_{nm} = \frac{1}{nm(m-1)} \sum_{i = 1}^n  \sum_{k=1}^m \sum_{\substack{l = 1 \\ l \neq k}}^m h_W \big( (X_i, R_i^X), (X_i, R_i^X); (Y_k, R_k^Y), (Y_l, R_l^Y)   \big) ,
\end{align*}
and
\begin{align*}
    Q_{nm} = \frac{1}{nm} \sum_{i = 1}^n  \sum_{k=1}^m  h_W \big( (X_i, R_i^X), (X_i, R_i^X); (Y_k, R_k^Y), (Y_k, R_k^Y)   \big).
\end{align*}

It is clear that, as $n,m \to \infty$, $n/(n+m) \to \lambda^2$, 
\begin{align}\label{nm_convergences}
    \left(\frac{n}{ n+m} - \frac{1}{n+m} \right) \to \lambda^2, \quad \left(\frac{m}{ n+m} - \frac{1}{n+m} \right) \to 1 - \lambda^2, \quad \frac{1}{n+m} \to 0.
\end{align}
Next, by \eqref{neuhaus_convergence}, we have that
\begin{align}\label{M_convergence}
    M_{nm} \overset{D}{\to} U(1,1).
\end{align}
By the Law of Large Numbers for $U$-statistics, we have that
\begin{align}\label{N_convergence}
    N_{nm} \overset{P}{\to} \E \left( h_W \left((X_1, R_1^X), (X_2, R_2^X); (Y_1, R_1^Y), (Y_1, R_1^Y) \right) \right),
\end{align}
\begin{align}\label{P_convergence}
    P_{nm} \overset{P}{\to} \E \left( h_W \left((X_1, R_1^X), (X_1, R_1^X); (Y_1, R_1^Y), (Y_2, R_2^Y) \right) \right),
\end{align}
and 
\begin{align}\label{Q_convergence}
    Q_{nm} \overset{P}{\to} \E \left(h_W \left( (X_1, R_1^X), (X_1, R_1^X); (Y_1, R_1^Y), (Y_1, R_1^Y) \right) \right).
\end{align}

Combining \eqref{nm_convergences}, \eqref{M_convergence}, \eqref{N_convergence}, \eqref{P_convergence}, and \eqref{Q_convergence}, we conclude that
\begin{align*}
    \frac{nm}{n+m}T_{nm}^W & \overset{D}{\to} U(1,1) + \lambda^2 \E \left( h_W \left((X_1, R_1^X), (X_2, R_2^X); (Y_1, R_1^Y), (Y_1, R_1^Y) \right) \right) \nonumber \\
    & \hspace{40pt}+ (1 - \lambda^2) \E \left( h_W \left((X_1, R_1^X), (X_1, R_1^X); (Y_1, R_1^Y), (Y_2, R_2^Y) \right) \right) \\
    &= U(1,1) + \E \left( h_W \left((X_1, R_1^X), (X_2, R_2^X); (Y_1, R_1^Y), (Y_1, R_1^Y) \right) \right),
\end{align*}
where the last equality is due to the symmetry of $h_W$ and due to the assumption that $(X, R_X) \overset{D}{=} (Y, R_Y)$. Noting that
\begin{align*}
   \E \left( h_W \left((X_1, R_1^X), (X_2, R_2^X); (Y_1, R_1^Y), (Y_1, R_1^Y) \right) \right) =  \E \left[ \rho_W \left((X_1, R_1^X), (Y_1, R_1^Y)\right)  \right]
\end{align*}
concludes the proof of the theorem.

\newpage
\section{Simulation results}\label{appendix_results}

\begin{table}[ht] 
\caption{The legend for the tables presenting results for MCAR, MAR 1 to 9, and MAR rank mechanisms} \vspace{5pt}
\small
    \centering
\begin{tabular}{l|ll}
    \toprule
       \%  &  \multicolumn{2}{c}{Distribution of $Y$} \\
         \midrule
        \multirow{4}{*}{Distribution of $X$} & Algorithm \ref{algorithm_preserving} with $T_{nm}^{CC}$ & Algorithm \ref{algorithm_imputation} with mean imputation \\
        & Algorithm \ref{algorithm_pooled} with $T_{nm}^{CC}$ & Algorithm \ref{algorithm_imputation} with median imputation \\
        & Algorithm \ref{algorithm_preserving} with $T_{nm}^{W}$ & Algorithm \ref{algorithm_imputation} with 6NN imputation \\
        & Algorithm \ref{algorithm_pooled} with $T_{nm}^{W}$ & Algorithm \ref{algorithm_imputation} with missForest imputation \\
        \hline
    \end{tabular}
    \label{tab:energy:legend}
\end{table}

\begin{table}[ht] 
\caption{The legend for the tables presenting results for MAR logistic mechanism} \vspace{5pt}
\small
    \centering
    \begin{tabular}{l|ll}
    \toprule
       \%  &  \multicolumn{2}{c}{Distribution of $Y$} \\
         \midrule
        \multirow{5}{*}{Distribution of $X$} &  \makecell[vl]{Average missingness rate per  \\ \underline{incomplete variable of $x$ (in \%)} }& \makecell[vl]{Average missingness rate per  \\ \underline{incomplete variable of $y$ (in \%)} }  \\  
        & Algorithm \ref{algorithm_preserving} with $T_{nm}^{CC}$ & Algorithm \ref{algorithm_imputation} with mean imputation \\
        & Algorithm \ref{algorithm_pooled} with $T_{nm}^{CC}$ & Algorithm \ref{algorithm_imputation} with median imputation \\
        & Algorithm \ref{algorithm_preserving} with $T_{nm}^{W}$ & Algorithm \ref{algorithm_imputation} with 6NN imputation \\
        & Algorithm \ref{algorithm_pooled} with $T_{nm}^{W}$ & Algorithm \ref{algorithm_imputation} with missForest imputation \\
        \hline
    \end{tabular}
    \label{tab:energy:legend_logistic}
\end{table}

\begin{table}[ht]
\caption{Percentage of rejections (rounded to the nearest integer) for trivariate data missing according to the MCAR mechanism, $p = (0.1, 0.1, 0.1)$, $n=100$, $m = 50$, $N = 5000$ } \vspace{5pt}
\centering
\label{tab:energy:powers_MCAR_01}
\scriptsize
\begin{tabular}{l|ll|ll|ll|ll|ll|ll}
\%                                        & \multicolumn{2}{l|}{$\mathcal{N} (0, I)$} & \multicolumn{2}{l|}{$\mathcal{N} (0, C_1)$} & \multicolumn{2}{l|}{$\mathcal{N} (m_1, C_1)$} & \multicolumn{2}{l|}{$\mathcal{N} (0, C_2)$} & \multicolumn{2}{l|}{$\mathcal{N} (m_1, C_2)$} & \multicolumn{2}{l}{$t_5 (0, I)$} \\ \hline
\multirow{4}{*}{$\mathcal{N} (0, I)$}     & 4                   & 4                  & 25                   & 25                  & 100                   & 100                  & 13                  & 11                   & 92                    & 96                   & 12              & 9              \\
                                          & 4                   & 4                  & 24                   & 24                  & 100                   & 100                  & 11                  & 11                   & 91                    & 96                   & 12              & 9              \\
                                          & 5                   & 5                  & 40                   & 14                  & 100                   & 97                   & 14                  & 8                    & 96                    & 81                   & 13              & 9              \\
                                          & 5                   & 4                  & 38                   & 26                  & 100                   & 100                  & 14                  & 15                   & 96                    & 96                   & 14              & 10             \\ \hline 
\multirow{4}{*}{$\mathcal{N} (0, C_1)$}   & 47                  & 46                 & 5                    & 5                   & 100                   & 100                  & 53                  & 53                   & 99                    & 100                  & 77              & 80             \\
\hline             
\end{tabular}
\end{table}

\begin{table}[ht]
\caption{Percentage of rejections (rounded to the nearest integer) for trivariate data missing according to the MCAR mechanism, $p = (0.4, 0.4, 0.4)$, $n=100$, $m = 50$, $N = 5000$ } \vspace{5pt}
\centering
\label{tab:energy:powers_MCAR_04}
\scriptsize 
\begin{tabular}{l|ll|ll|ll|ll|ll|ll}
\%                                        & \multicolumn{2}{l|}{$\mathcal{N} (0, I)$} & \multicolumn{2}{l|}{$\mathcal{N} (0, C_1)$} & \multicolumn{2}{l|}{$\mathcal{N} (m_1, C_1)$} & \multicolumn{2}{l|}{$\mathcal{N} (0, C_2)$} & \multicolumn{2}{l|}{$\mathcal{N} (m_1, C_2)$} & \multicolumn{2}{l}{$t_5 (0, I)$}  \\
\hline
\multirow{4}{*}{$\mathcal{N} (0, I)$}     & 5                   & 5                  & 4                   & 6                    & 5                    & 98                    & 6                    & 6                   & 35                    & 86                   & 9               & 9              \\
                                          & 5                   & 5                  & 4                   & 6                    & 5                    & 95                    & 6                    & 5                   & 33                    & 80                   & 9               & 7              \\
                                          & 5                   & 4                  & 22                  & 3                    & 10                   & 23                    & 8                    & 5                   & 79                    & 20                   & 11              & 8              \\
                                          & 5                   & 5                  & 20                  & 7                    & 10                   & 94                    & 7                    & 8                   & 79                    & 78                   & 10              & 9              \\ \hline 
\multirow{4}{*}{$\mathcal{N} (0, C_1)$}   & 14                  & 15                 & 5                   & 5                    & 63                   & 100                   & 15                   & 14                  & 54                    & 95                   & 24              & 26             \\
\hline 
\end{tabular}
\end{table}

\begin{table}[ht]
\caption{Percentage of rejections (rounded to the nearest integer) for trivariate data missing according to the MCAR mechanism, $p = (0.1, 0.2, 0.3)$, $n=100$, $m = 50$, $N = 5000$ } \vspace{5pt}
\centering
\label{tab:energy:powers_MCAR_01_02_03}
\scriptsize 
\begin{tabular}{l|ll|ll|ll|ll|ll|ll}
\%                                        & \multicolumn{2}{l|}{$\mathcal{N} (0, I)$} & \multicolumn{2}{l|}{$\mathcal{N} (0, C_1)$} & \multicolumn{2}{l|}{$\mathcal{N} (m_1, C_1)$} & \multicolumn{2}{l|}{$\mathcal{N} (0, C_2)$} & \multicolumn{2}{l|}{$\mathcal{N} (m_1, C_2)$} & \multicolumn{2}{l}{$t_5 (0, I)$}  \\
\hline
\multirow{4}{*}{$\mathcal{N} (0, I)$}     & 5                  & 5                   & 12                   & 15                  & 96                    & 100                  & 9                    & 7                   & 78                    & 93                   & 8               & 8              \\
                                          & 5                  & 5                   & 12                   & 14                  & 95                    & 100                  & 9                    & 7                   & 77                    & 91                   & 9               & 7              \\
                                          & 5                  & 5                   & 31                   & 6                   & 99                    & 53                   & 11                   & 6                   & 92                    & 39                   & 11              & 8              \\
                                          & 5                  & 4                   & 32                   & 17                  & 99                    & 100                  & 11                   & 11                  & 92                    & 94                   & 11              & 9              \\ \hline 
\multirow{4}{*}{$\mathcal{N} (0, C_1)$}   & 30                 & 31                  & 5                    & 5                   & 99                    & 100                  & 35                   & 31                  & 92                    & 99                   & 59              & 60             \\
\hline            
\end{tabular}
\end{table}

\begin{table}[ht]
\caption{Percentage of rejections (rounded to the nearest integer) for trivariate data missing according to the MAR 1 to 9 mechanism, $p = (0, 0.1, 0.1)$ (Var. 1 controls missingness in Var. 2 and Var. 3), $n=100$, $m = 50$, $N = 5000$ } \vspace{5pt}
\scriptsize 
\label{tab:energy:powers_MAR1to9_01}
\centering
\begin{tabular}{l|ll|ll|ll|ll|ll|ll}
\%                                        & \multicolumn{2}{l|}{$\mathcal{N} (0, I)$} & \multicolumn{2}{l|}{$\mathcal{N} (0, C_1)$} & \multicolumn{2}{l|}{$\mathcal{N} (m_1, C_1)$} & \multicolumn{2}{l|}{$\mathcal{N} (0, C_2)$} & \multicolumn{2}{l|}{$\mathcal{N} (m_1, C_2)$} & \multicolumn{2}{l}{$t_5 (0, I)$}  \\
\hline
\multirow{4}{*}{$\mathcal{N} (0, I)$}     & 3                   & 3                  & 29                   & 28                  & 100                   & 100                  & 14                   & 8                   & 90                    & 96                   & 9               & 8              \\
                                          & 3                   & 3                  & 28                   & 27                  & 100                   & 100                  & 14                   & 8                   & 90                    & 96                   & 9               & 7              \\
                                          & 4                   & 3                  & 42                   & 16                  & 100                   & 100                  & 15                   & 7                   & 96                    & 83                   & 10              & 9              \\
                                          & 4                   & 3                  & 43                   & 26                  & 100                   & 100                  & 16                   & 10                  & 97                    & 98                   & 13              & 7              \\ \hline 
\multirow{4}{*}{$\mathcal{N} (0, C_1)$}   & 52                  & 49                 & 3                    & 3                   & 100                   & 100                  & 63                   & 54                  & 98                    & 100                  & 86              & 84             \\
\hline      
\end{tabular}
\end{table}

\begin{table}[ht]
\caption{Percentage of rejections (rounded to the nearest integer) for trivariate data missing according to the MAR 1 to 9 mechanism, $p = (0, 0.4, 0.4)$ (Var. 1 controls missingness in Var. 2 and Var. 3), $n=100$, $m = 50$, $N = 5000$ } \vspace{5pt}
\scriptsize
\label{tab:energy:powers_MAR1to9_04}
\centering
\begin{tabular}{l|ll|ll|ll|ll|ll|ll}
\%                                        & \multicolumn{2}{l|}{$\mathcal{N} (0, I)$} & \multicolumn{2}{l|}{$\mathcal{N} (0, C_1)$} & \multicolumn{2}{l|}{$\mathcal{N} (m_1, C_1)$} & \multicolumn{2}{l|}{$\mathcal{N} (0, C_2)$} & \multicolumn{2}{l|}{$\mathcal{N} (m_1, C_2)$} & \multicolumn{2}{l}{$t_5 (0, I)$} \\
\hline
\multirow{4}{*}{$\mathcal{N} (0, I)$}     & 6                   & 5                  & 25                   & 7                   & 100                   & 99                   & 39                   & 16                  & 55                    & 63                   & 11              & 7              \\
                                          & 5                   & 5                  & 22                   & 7                   & 99                    & 98                   & 36                   & 14                  & 55                    & 57                   & 13              & 6              \\
                                          & 9                   & 6                  & 48                   & 5                   & 100                   & 27                   & 37                   & 7                   & 89                    & 14                   & 18              & 9              \\
                                          & 9                   & 4                  & 43                   & 7                   & 100                   & 94                   & 36                   & 6                   & 88                    & 74                   & 18              & 7              \\ \hline 
\multirow{4}{*}{$\mathcal{N} (0, C_1)$}   & 41                  & 20                 & 6                    & 5                   & 99                    & 100                  & 67                   & 39                  & 59                    & 91                   & 65              & 37             \\
\hline 
\end{tabular}
\end{table}

\begin{table}[ht]
\caption{Percentage of rejections (rounded to the nearest integer) for trivariate data missing according to the MAR rank mechanism, $p = (0, 0.1, 0.1)$ (Var. 1 controls missingness in Var. 2 and Var. 3), $n=100$, $m = 50$, $N = 5000$ } \vspace{5pt}
\scriptsize
\label{tab:energy:powers_MARrank_01}
\centering
\begin{tabular}{l|ll|ll|ll|ll|ll|ll}
\%                                        & \multicolumn{2}{l|}{$\mathcal{N} (0, I)$} & \multicolumn{2}{l|}{$\mathcal{N} (0, C_1)$} & \multicolumn{2}{l|}{$\mathcal{N} (m_1, C_1)$} & \multicolumn{2}{l|}{$\mathcal{N} (0, C_2)$} & \multicolumn{2}{l|}{$\mathcal{N} (m_1, C_2)$} & \multicolumn{2}{l}{$t_5 (0, I)$}  \\
\hline
\multirow{4}{*}{$\mathcal{N} (0, I)$}     & 4                   & 3                  & 28                   & 28                  & 100                   & 100                  & 12                   & 10                  & 92                    & 97                   & 10              & 8              \\
                                          & 4                   & 3                  & 28                   & 28                  & 100                   & 100                  & 12                   & 9                   & 92                    & 97                   & 9               & 8              \\
                                          & 4                   & 5                  & 40                   & 17                  & 100                   & 100                  & 12                   & 8                   & 96                    & 85                   & 12              & 8              \\
                                          & 3                   & 3                  & 40                   & 29                  & 100                   & 100                  & 12                   & 11                  & 96                    & 98                   & 13              & 8              \\ \hline 
\multirow{4}{*}{$\mathcal{N} (0, C_1)$}   & 51                  & 49                 & 4                    & 3                   & 100                   & 100                  & 61                   & 57                  & 99                    & 100                  & 84              & 84             \\
\hline 
\end{tabular}
\end{table}

\begin{table}[ht]
\caption{Percentage of rejections (rounded to the nearest integer) for trivariate data missing according to the MAR rank mechanism, $p = (0, 0.4, 0.4)$ (Var. 1 controls missingness in Var. 2 and Var. 3), $n=100$, $m = 50$, $N = 5000$ } \vspace{5pt}
\scriptsize
\label{tab:energy:powers_MARrank_04}
\centering
\begin{tabular}{l|ll|ll|ll|ll|ll|ll}
\%                                        & \multicolumn{2}{l|}{$\mathcal{N} (0, I)$} & \multicolumn{2}{l|}{$\mathcal{N} (0, C_1)$} & \multicolumn{2}{l|}{$\mathcal{N} (m_1, C_1)$} & \multicolumn{2}{l|}{$\mathcal{N} (0, C_2)$} & \multicolumn{2}{l|}{$\mathcal{N} (m_1, C_2)$} & \multicolumn{2}{l}{$t_5 (0, I)$}  \\
\hline
\multirow{4}{*}{$\mathcal{N} (0, I)$}     & 5                   & 4                  & 16                   & 6                   & 98                    & 99                   & 31                   & 14                  & 43                    & 69                   & 12              & 6              \\
                                          & 4                   & 4                  & 15                   & 6                   & 98                    & 98                   & 29                   & 12                  & 43                    & 62                   & 12              & 6              \\
                                          & 5                   & 5                  & 36                   & 4                   & 99                    & 24                   & 23                   & 5                   & 83                    & 14                   & 14              & 9              \\
                                          & 5                   & 4                  & 36                   & 7                   & 99                    & 97                   & 22                   & 6                   & 82                    & 78                   & 14              & 6              \\ \hline 
\multirow{4}{*}{$\mathcal{N} (0, C_1)$}   & 33                  & 18                 & 5                    & 4                   & 97                    & 100                  & 56                   & 36                  & 51                    & 92                   & 60              & 37             \\
\hline 
\end{tabular}
\end{table}

\begin{table}[ht]
\caption{Percentage of rejections (rounded to the nearest integer) for trivariate data missing according to the MAR logistic mechanism, for intercept equal to $-5$ and slope parameters $-1.9$ and $-1.5$ (Var. 1 controls missingness in Var. 2 and Var. 3), $n=100$, $m = 50$, $N = 5000$ } \vspace{5pt}
\scriptsize 
\label{tab:energy:powers_MAR_logistic_3D_m6_m1p9_m1p5}
\centering
\begin{tabular}{l|ll|ll|ll|ll|ll|ll}
\%                                        & \multicolumn{2}{l|}{$\mathcal{N} (0, I)$} & \multicolumn{2}{l|}{$\mathcal{N} (0, C_1)$} & \multicolumn{2}{l|}{$\mathcal{N} (m_1, C_1)$} & \multicolumn{2}{l|}{$\mathcal{N} (0, C_2)$} & \multicolumn{2}{l|}{$\mathcal{N} (m_1, C_2)$} & \multicolumn{2}{l}{$t_5 (0, I)$}  \\
\hline
\multirow{5}{*}{$\mathcal{N} (0, I)$}     & \underline{10}                  & \underline{10}                 & \underline{10}                   & \underline{5}                   & \underline{10}                    & \underline{1}                    & \underline{10}                   & \underline{10}                  & \underline{10}                    & \underline{4}                    & \underline{10}              & \underline{12}             \\
                                          & 5                   & 4                  & 40                   & 30                  & 100                   & 100                  & 18                   & 12                  & 98                    & 98                   & 10              & 10             \\
                                          & 5                   & 4                  & 38                   & 29                  & 100                   & 100                  & 16                   & 12                  & 98                    & 98                   & 10              & 10             \\
                                          & 4                   & 5                  & 64                   & 20                  & 100                   & 100                  & 16                   & 8                   & 98                    & 95                   & 14              & 12             \\
\hline            
\end{tabular}
\end{table}

\begin{table}[ht]
\caption{Percentage of rejections (rounded to the nearest integer) for trivariate data missing according to the MAR logistic mechanism, for intercept equal to $-1.05$ and slope parameters $-1.7$ and $-0.6$ (Var. 1 controls missingness in Var. 2 and Var. 3), $n=100$, $m = 50$, $N = 5000$ } \vspace{5pt}
\scriptsize 
\label{tab:energy:powers_MAR_logistic_3D_m1p05_m1p7_m0p6}
\centering
\begin{tabular}{l|ll|ll|ll|ll|ll|ll}
\%                                        & \multicolumn{2}{l|}{$\mathcal{N} (0, I)$} & \multicolumn{2}{l|}{$\mathcal{N} (0, C_1)$} & \multicolumn{2}{l|}{$\mathcal{N} (m_1, C_1)$} & \multicolumn{2}{l|}{$\mathcal{N} (0, C_2)$} & \multicolumn{2}{l|}{$\mathcal{N} (m_1, C_2)$} & \multicolumn{2}{l}{$t_5 (0, I)$} \\ \hline 
\multirow{5}{*}{$\mathcal{N} (0, I)$}     & \underline{36}                  & \underline{36}                 &\underline{36}                   & \underline{33}                  & \underline{36}                    & \underline{18}                   & \underline{36}                   & \underline{36}                  & \underline{36}                    & \underline{22}                   & \underline{36}              & \underline{37}             \\
                                          & 5                   & 5                  & 28                   & 10                  & 96                    & 100                  & 38                   & 21                  & 98                    & 98                   & 12              & 8              \\
                                          & 5                   & 5                  & 27                   & 9                   & 96                    & 100                  & 37                   & 18                  & 97                    & 97                   & 11              & 7              \\
                                          & 5                   & 5                  & 50                   & 4                   & 100                   & 42                   & 28                   & 7                   & 97                    & 37                   & 16              & 8              \\
\hline              
\end{tabular}
\end{table}

\begin{table}[ht]
\caption{Percentage of rejections (rounded to the nearest integer) for decavariate data missing according to the MCAR mechanism, $p = (0.05, \dots, 0.05)$, $n=100$, $m = 50$, $N = 5000$ } \vspace{5pt}
\centering
\label{tab:energy:10D_powers_MCAR_005}
\scriptsize
\begin{tabular}{l|ll|ll|ll|ll|ll|ll}
\%                                        & \multicolumn{2}{l|}{$\mathcal{N} (0, I)$} & \multicolumn{2}{l|}{$\mathcal{N} (0, C_1)$} & \multicolumn{2}{l|}{$\mathcal{N} (m_1, C_1)$} & \multicolumn{2}{l|}{$\mathcal{N} (0, C_2)$} & \multicolumn{2}{l|}{$\mathcal{N} (m_1, C_2)$} & \multicolumn{2}{l}{$t_5 (0, I)$}  \\
\hline
\multirow{4}{*}{$\mathcal{N} (0, I)$}     & 5                   & 4                  & 41                   & 86                  & 100                   & 100                  & 23                   & 36                  & 97                    & 100                  & 17              & 21             \\
                                          & 5                   & 4                  & 40                   & 86                  & 100                   & 100                  & 24                   & 36                  & 97                    & 100                  & 16              & 21             \\
                                          & 5                   & 5                  & 83                   & 72                  & 100                   & 100                  & 51                   & 32                  & 100                   & 100                  & 24              & 20             \\
                                          & 5                   & 5                  & 84                   & 86                  & 100                   & 100                  & 52                   & 50                  & 100                   & 100                  & 24              & 21             \\ \hline 
\multirow{4}{*}{$\mathcal{N} (0, C_1)$}   & 78                  & 97                 & 5                    & 4                   & 100                   & 100                  & 76                   & 93                  & 100                   & 100                  & 95              & 100            \\
\hline         
\end{tabular}
\end{table}

\begin{table}[ht]
\caption{Percentage of rejections (rounded to the nearest integer) for decavariate data missing according to the MAR 1 to 9 mechanism, $p = (0,0,0,0,0, 0.05, \dots, 0.05)$ (Vars. 1--5 control missingness in Vars. 6--10), $n=100$, $m = 50$, $N = 5000$ } \vspace{5pt}
\centering
\label{tab:energy:10D_powers_MAR_1to9_005}
\scriptsize
\begin{tabular}{l|ll|ll|ll|ll|ll|ll}
\%                                        & \multicolumn{2}{l|}{$\mathcal{N} (0, I)$} & \multicolumn{2}{l|}{$\mathcal{N} (0, C_1)$} & \multicolumn{2}{l|}{$\mathcal{N} (m_1, C_1)$} & \multicolumn{2}{l|}{$\mathcal{N} (0, C_2)$} & \multicolumn{2}{l|}{$\mathcal{N} (m_1, C_2)$} & \multicolumn{2}{l}{$t_5 (0, I)$}  \\
\hline
\multirow{4}{*}{$\mathcal{N} (0, I)$}     & 5                   & 4                  & 75                   & 89                  & 100                   & 100                  & 43                   & 46                  & 99                    & 100                  & 22              & 21             \\
                                          & 5                   & 4                  & 77                   & 89                  & 100                   & 100                  & 42                   & 46                  & 99                    & 100                  & 22              & 21             \\
                                          & 6                   & 5                  & 95                   & 83                  & 100                   & 100                  & 59                   & 41                  & 100                   & 100                  & 22              & 24             \\
                                          & 7                   & 6                  & 96                   & 89                  & 100                   & 100                  & 57                   & 51                  & 100                   & 100                  & 23              & 26             \\ \hline 
\multirow{4}{*}{$\mathcal{N} (0, C_1)$}   & 95                  & 99                 & 5                    & 4                   & 100                   & 100                  & 94                   & 95                  & 100                   & 100                  & 99              & 100            \\
\hline           
\end{tabular}
\end{table}

\begin{table}[ht]
\caption{Percentage of rejections (rounded to the nearest integer) for decavariate data missing according to the MAR rank mechanism, $p = (0,0,0,0,0, 0.05, \dots, 0.05)$ (Vars. 1--5 control missingness in Vars. 6--10), $n=100$, $m = 50$, $N = 5000$ } \vspace{5pt}
\centering
\label{tab:energy:10D_powers_MAR_rank_005}
\scriptsize
\begin{tabular}{l|ll|ll|ll|ll|ll|ll}
\%                                        & \multicolumn{2}{l|}{$\mathcal{N} (0, I)$} & \multicolumn{2}{l|}{$\mathcal{N} (0, C_1)$} & \multicolumn{2}{l|}{$\mathcal{N} (m_1, C_1)$} & \multicolumn{2}{l|}{$\mathcal{N} (0, C_2)$} & \multicolumn{2}{l|}{$\mathcal{N} (m_1, C_2)$} & \multicolumn{2}{l}{$t_5 (0, I)$} \\
\hline
\multirow{4}{*}{$\mathcal{N} (0, I)$}     & 5                   & 4                  & 71                   & 91                  & 100                   & 100                  & 40                   & 46                  & 98                    & 100                  & 20              & 24             \\
                                          & 5                   & 4                  & 72                   & 91                  & 100                   & 100                  & 40                   & 46                  & 99                    & 100                  & 20              & 24             \\
                                          & 5                   & 5                  & 88                   & 83                  & 100                   & 100                  & 53                   & 40                  & 100                   & 100                  & 23              & 22             \\
                                          & 5                   & 5                  & 89                   & 90                  & 100                   & 100                  & 53                   & 54                  & 100                   & 100                  & 22              & 23             \\ \hline 
\multirow{4}{*}{$\mathcal{N} (0, C_1)$}   & 94                  & 98                 & 5                    & 4                   & 100                   & 100                  & 93                   & 95                  & 100                   & 100                  & 99              & 100            \\
\hline             
\end{tabular}
\end{table}

\begin{table}[ht]
\caption{Percentage of rejections (rounded to the nearest integer) for decavariate data missing according to the MAR logistic mechanism, for intercept equal to $-2.6$ and slope parameters $0.3$, $1.8$, $1.7$, $-1.2$ and $1.9$ (Vars. 1--5 control missingness in Vars. 6--10), $n=100$, $m = 50$, $N = 5000$ }\vspace{5pt}
\centering
\label{tab:energy:10D_powers_MAR_logistic}
\scriptsize
\begin{tabular}{l|ll|ll|ll|ll|ll|ll}
\%                                        & \multicolumn{2}{l|}{$\mathcal{N} (0, I)$} & \multicolumn{2}{l|}{$\mathcal{N} (0, C_1)$} & \multicolumn{2}{l|}{$\mathcal{N} (m_1, C_1)$} & \multicolumn{2}{l|}{$\mathcal{N} (0, C_2)$} & \multicolumn{2}{l|}{$\mathcal{N} (m_1, C_2)$} & \multicolumn{2}{l}{$t_5 (0, I)$} \\
\hline
\multirow{5}{*}{$\mathcal{N} (0, I)$}     & \underline{25}                  & \underline{25}                 & \underline{25}                   & \underline{19}                  & \underline{25}                    & \underline{45}                   & \underline{25}                   & \underline{28}                  & \underline{25}                   & \underline{47}                    & \underline{25}              & \underline{26}             \\
                                          & 4                   & 5                  & 46                   & 62                  & 98                    & 100                  & 86                   & 57                  & 9                    & 98                    & 17              & 18             \\
                                          & 5                   & 5                  & 50                   & 60                  & 98                    & 100                  & 86                   & 55                  & 9                    & 98                    & 18              & 17             \\
                                          & 5                   & 5                  & 93                   & 12                  & 95                    & 89                   & 63                   & 15                  & 47                   & 49                    & 25              & 11             \\
\hline             
\end{tabular}
\end{table}


\begin{table}[ht]
\caption{Percentage of rejections (rounded to the nearest integer) for\texttt{wine\_quality} data missing according to the MCAR mechanism, $p = (0.1, 0.1, 0.1)$ (left), $p = (0.1, 0.2, 0.3)$ (middle) and $p = (0.4, 0.4, 0.4)$ (right), $n=100$, $m = 50$, $N = 5000$ } \vspace{10pt}
\centering
\label{tab:energy:winequality_powers_MCAR_01_04}
\scriptsize
\begin{minipage}{0.33\textwidth}\centering
    \begin{tabular}{l|ll|ll}
\%                     & \multicolumn{2}{l|}{White} & \multicolumn{2}{l}{Red} \\ \hline 
\multirow{4}{*}{White} & 5           & 5           & 23         & 25         \\
                       & 5           & 5           & 22         & 22         \\
                       & 5           & 5           & 41         & 22         \\
                       & 6           & 5           & 44         & 26         \\ \hline 
\multirow{4}{*}{Red}   & 29          & 37          & 5          & 5          \\
                       & 28          & 34          & 6          & 5          \\
                       & 48          & 31          & 5          & 5          \\
                       & 48          & 39          & 5          & 5  \\ \hline         
\end{tabular}
\end{minipage}%
\begin{minipage}{0.33\textwidth} \centering
    \begin{tabular}{l|ll|ll}
\%                     & \multicolumn{2}{l|}{White} & \multicolumn{2}{l}{Red} \\ \hline 
\multirow{4}{*}{White} & 6           & 5           & 13         & 14         \\
                       & 6           & 5           & 13         & 10         \\
                       & 6           & 5           & 59         & 8          \\
                       & 5           & 5           & 62         & 15         \\ \hline 
\multirow{4}{*}{Red}   & 19          & 19          & 6          & 5          \\
                       & 16          & 15          & 5          & 5          \\
                       & 65          & 11          & 6          & 5          \\
                       & 63          & 19          & 5          & 5            \\ \hline          
\end{tabular}
\end{minipage}%
\begin{minipage}{0.33\textwidth} \centering
    \begin{tabular}{l|ll|ll}
\%                     & \multicolumn{2}{l|}{White} & \multicolumn{2}{l}{Red} \\ \hline 
\multirow{4}{*}{White} & 5           & 5           & 6          & 11         \\
                       & 5           & 5           & 6          & 8          \\
                       & 5           & 5           & 36         & 4          \\
                       & 5           & 5           & 36         & 10         \\ \hline 
\multirow{4}{*}{Red}   & 10          & 17          & 5          & 5          \\
                       & 9           & 15          & 6          & 5          \\
                       & 44          & 9           & 5          & 5          \\
                       & 45          & 16          & 5          & 5     \\ \hline          
\end{tabular}
\end{minipage}

\end{table}

\begin{table}[ht]
\caption{Percentage of rejections (rounded to the nearest integer) for \texttt{wine\_quality} data missing according to the MAR 1 to 9 mechanism, $p = (0, 0.1, 0.1)$ (left) and $p = (0, 0.4, 0.4)$ (right), $n=100$, $m = 50$, $N = 5000$ } \vspace{10pt}
\centering
\label{tab:energy:winequality_powers_MAR1to9_01_04}
\scriptsize
\begin{minipage}{0.5\textwidth}\centering
    \begin{tabular}{l|ll|ll}
\%                     & \multicolumn{2}{l|}{White} & \multicolumn{2}{l}{Red} \\ \hline 
\multirow{4}{*}{White} & 5           & 5           & 27         & 27         \\
                       & 5           & 5           & 28         & 23         \\
                       & 5           & 5           & 44         & 27         \\
                       & 6           & 6           & 52         & 29         \\ \hline 
\multirow{4}{*}{Red}   & 36          & 34          & 5          & 5          \\
                       & 34          & 29          & 5          & 5          \\
                       & 56          & 31          & 6          & 4          \\
                       & 58          & 36          & 6          & 5     \\ \hline         
\end{tabular}
\end{minipage}%
\begin{minipage}{0.5\textwidth} \centering
    \begin{tabular}{l|ll|ll}
\%                     & \multicolumn{2}{l|}{White} & \multicolumn{2}{l}{Red} \\ \hline 
\multirow{4}{*}{White} & 6           & 5           & 14         & 11         \\
                       & 6           & 5           & 15         & 8          \\
                       & 6           & 5           & 63         & 6          \\
                       & 7           & 5           & 73         & 12         \\ \hline 
\multirow{4}{*}{Red}   & 20          & 16          & 5          & 4          \\
                       & 22          & 14          & 5          & 4          \\
                       & 73          & 11          & 6          & 5          \\
                       & 83          & 15          & 6          & 4    \\ \hline          
\end{tabular}
\end{minipage}
\end{table}

\begin{table}[ht]
\caption{Percentage of rejections (rounded to the nearest integer) for \texttt{wine\_quality} data missing according to the MAR rank mechanism, $p = (0, 0.1, 0.1)$ (left) and $p = (0, 0.4, 0.4)$ (right), $n=100$, $m = 50$, $N = 5000$ } \vspace{10pt}
\centering
\label{tab:energy:winequality_powers_MARrank_01_04}
\scriptsize
\begin{minipage}{0.5\textwidth}\centering
    \begin{tabular}{l|ll|ll}
\%                     & \multicolumn{2}{l|}{White} & \multicolumn{2}{l}{Red} \\ \hline 
\multirow{4}{*}{White} & 5           & 5           & 25         & 24         \\
                       & 5           & 5           & 29         & 20         \\
                       & 5           & 5           & 43         & 22         \\
                       & 5           & 5           & 44         & 25         \\ \hline 
\multirow{4}{*}{Red}   & 39          & 34          & 6          & 6          \\
                       & 35          & 28          & 5          & 5          \\
                       & 55          & 30          & 6          & 6          \\
                       & 55          & 34          & 6          & 5      \\ \hline         
\end{tabular}
\end{minipage}%
\begin{minipage}{0.5\textwidth} \centering
    \begin{tabular}{l|ll|ll}
\%                     & \multicolumn{2}{l|}{White} & \multicolumn{2}{l}{Red} \\ \hline 
\multirow{4}{*}{White} & 5           & 5           & 14         & 12         \\
                       & 5           & 5           & 15         & 8          \\
                       & 5           & 5           & 60         & 6          \\
                       & 5           & 5           & 64         & 11         \\ \hline 
\multirow{4}{*}{Red}   & 17          & 15          & 4          & 5          \\
                       & 16          & 14          & 5          & 5          \\
                       & 68          & 9           & 5          & 5          \\
                       & 68          & 14          & 5          & 5      \\ \hline          
\end{tabular}
\end{minipage}
\end{table}

\begin{table}[ht]
\caption{Percentage of rejections (rounded to the nearest integer) for \texttt{wine\_quality} data missing according to the MAR logistic mechanism; left: intercept equal to $-1.3$, slope parameters $0.7$ and $1.1$; right: intercept equal to $9$, slope parameters $-2$ and $-1$; $n=100$, $m = 50$, $N = 5000$ } \vspace{10pt}
\centering
\label{tab:energy:winequality_powers_MARlogistic}
\scriptsize
\begin{minipage}{0.5\textwidth}\centering
    \begin{tabular}{l|ll|ll}
\%                     & \multicolumn{2}{l|}{White} & \multicolumn{2}{l}{Red} \\ \hline 
\multirow{5}{*}{White} & \underline{7}           & \underline{7}           & \underline{7}          & \underline{7}          \\
                       & 5           & 4           & 29         & 30         \\
                       & 5           & 4           & 26         & 27         \\
                       & 5           & 4           & 44         & 28         \\
                       & 5           & 4           & 45         & 29         \\ \hline 
\multirow{5}{*}{Red}   & \underline{7}           & \underline{7}           & \underline{7}          & \underline{7}          \\
                       & 32          & 35          & 5          & 6          \\
                       & 35          & 32          & 5          & 5          \\
                       & 47          & 33          & 6          & 5          \\
                       & 48          & 35          & 6          & 5       \\ \hline         
\end{tabular}
\end{minipage}%
\begin{minipage}{0.5\textwidth} \centering
    \begin{tabular}{l|ll|ll}
\%                     & \multicolumn{2}{l|}{White} & \multicolumn{2}{l}{Red} \\ \hline 
\multirow{5}{*}{White} & \underline{37}          & \underline{37}          & \underline{37}         & \underline{29}         \\
                       & 5           & 5           & 12         & 12         \\
                       & 5           & 5           & 11         & 9          \\
                       & 5           & 5           & 58         & 5          \\
                       & 5           & 5           & 51         & 10         \\ \hline 
\multirow{5}{*}{Red}   & \underline{29}          & \underline{37}          & \underline{29}         & \underline{29}         \\
                       & 14          & 18          & 5          & 6          \\
                       & 15          & 15          & 4          & 5          \\
                       & 58          & 14          & 4          & 5          \\
                       & 50          & 17          & 5          & 5      \\ \hline          
\end{tabular}
\end{minipage}
\end{table}

\end{document}